\documentclass[a4paper,11pt]{article}
\pdfoutput=1 

\usepackage{jheppub} 

\usepackage[T1]{fontenc} 

\usepackage{flexisym}
\usepackage{breqn}
\usepackage{amsmath}
\usepackage{amssymb,amsfonts}
\usepackage{graphicx}
\usepackage[utf8]{inputenc}
\usepackage{color}
\usepackage{algorithm}
\usepackage{algorithmic}
\usepackage{listings}
\usepackage{textcomp}
\usepackage{url}
\usepackage{diagbox}
\usepackage{subfigure}
\usepackage{mathtools}
\usepackage[section]{placeins}

\newcommand{\C}{\mathbb{C}}

\newcommand{\be}{\begin{equation}}
\newcommand{\ee}{\end{equation}}
\newcommand{\dd}{\mathrm{d}}
\newcommand{\bse}{\begin{subequations}}
\newcommand{\ese}{\end{subequations}}

\newcommand{\ii}{\mathrm{i}}
\newcommand{\e}{\mathrm{e}}

\newcommand{\bpm}{\begin{pmatrix}}
\newcommand{\epm}{\end{pmatrix}}
\newcommand{\bmm}{\begin{matrix}}
\newcommand{\emm}{\end{matrix}}

\newcommand{\x}{\times}

\allowdisplaybreaks

\makeatletter
\newcommand*{\Relbarfill@}{\arrowfill@\Relbar\Relbar\Relbar}
\newcommand*{\xeq}[2][]{\ext@arrow 0055\Relbarfill@{#1}{#2}}
\makeatother

\newcommand{\bra}[1]{\langle {#1}|}
\newcommand{\ket}[1]{| #1 \rangle}
\newcommand{\SLDC}{{\mathrm{SL}(2,\mathbb{C})}}

\newcommand{\half}{\frac{1}{2}}

\newcommand{\Su}{\mathrm{SU}(2)}

\usepackage[normalem]{ulem}
\renewcommand\sout{\bgroup\markoverwith{\textcolor{red}{\rule[0.5ex]{4pt}{0.8pt}}}\ULon}

\def\mb{\left(\begin{matrix}}
\def\me{\end{matrix}\right)}

\newcommand{\Psix}[3][1]{
\begin{tikzpicture}[scale=0.8]
\node[name=s, regular polygon, regular polygon sides=6, minimum size=1cm, outer sep=0pt ,draw] at (0,0) {}; 
\foreach \anchor/\x/\y /\xx/\yy /\b in
{corner 1/0.17/0.17*1.732/-0.11/0.18/1, corner 2/-0.17/0.17*1.732/0.07/0.18/2, corner 3/-0.34/0/-0.15/-0.18/3, corner 4/-0.17/-0.17*1.732/-0.22/-0.05/4, corner 5/0.17/-0.17*1.732/0.2/-0.05/5, corner 6/0.34/0/0.15/-0.18/6}
{
 \draw[shift=(s.\anchor)] (0,0) -- (\x,\y) node at(\xx,\yy) {$#2_{\text{\scalebox{0.7}{$\b$}}}$};
 \ifnum #1=1
 \draw[shift=(s.\anchor),<-,>=stealth', line width=0.01pt] (s.\anchor) -- (\x,\y);
 \fi
 }
\foreach \anchor/\xx/\yy /\a in
{side 1/0/-0.18/1, side 2/-0.18/0.05/2, side 3/0.15/0.05/3, side 4/0/-0.18/4, side 5/-0.18/0.05/5, side 6/0.15/0.05/6}
 \draw[shift=(s.\anchor)]  node at(\xx,\yy) {$#3_{\text{\scalebox{0.7}{$\a$}}}$};
\ifnum #1=1{
  \foreach \anchorr/\anchorf in
   {corner 1/corner 2, corner 2/corner 3, corner 3/corner 4, corner 4/corner 5, corner 5/corner 6, corner 6/corner 1}
   \draw[shift=(s.\anchorr), ->, >=stealth', line width=0.01pt]  (s.\anchorr) -- (s.\anchorf);}
 \else {
  \foreach \anchorb/\anchorw in
   {corner 1/corner 2, corner 3/corner 4, corner 5/corner 6} {
   \node[fill=black, circle, minimum size=0, inner sep=0, outer sep=0, draw] at(s.\anchorb) {};
   \node[fill=white, circle, minimum size=0, inner sep=0, outer sep=0, draw] at(s.\anchorw) {};}
}
\fi
\end{tikzpicture}
}

\usepackage{varioref}

\title{\boldmath A saddle-point finder and its application to the spin foam model}

\author[a,b,1]{Zichang Huang,\note{Corresponding author}}
\author[a,b]{Shan Huang}
\author[a,b,c,1]{Yidun Wan}
 \affiliation[a]{State Key Laboratory of Surface Physics, Department of Physics, Center for Field Theory and Particle Physics, and Institute for Nanoelectronic devices and Quantum computing, Fudan University, Shanghai 200433, China}
 \affiliation[b]{Shanghai Qi Zhi Institute, Shanghai 200030, China}
 \affiliation[c]{Zhangjiang Fudan International Innovation Center, Fudan University, Shanghai 201210, China}

\emailAdd{hzc881126@hotmail.com, ydwan@fudan.edu.cn, shanhuang96@hotmail.com}

\abstract{We introduce a saddle-point finder that can find the complex saddle points for any analytically continued action. We showcase our saddle-point finder by two examples in the EPRL spin foam model: the single vertex case and the case of triangulation $\Delta_3$. In both cases, the complex saddle points are found, and each saddle point's contribution to the partition function is estimated. We also discuss the geometrical interpretation of each saddle point.}

\begin{document} 
\maketitle
\flushbottom

\section{Introduction}
Witten has suggested to use complex path integral to study the physical theories with complex-valued couplings \cite{Witten:2010cx,Harlow2011,Witten:2010zr}. 
Later, refs.\cite{BALIAN1974514,PhysRevD.15.1558,LAPEDES198258,BALITSKY1986475,Cristoforetti:2012su,Fujii:2013sra,Aarts:2014nxa,Fujii2015,Tanizaki_2016} have related complex path integrals to the sign problem in the Euclidean path integral of QCD and models with finite chemical potential.
Works on super-symmetric theories \cite{PhysRevD.80.065001,Poppitz2012,Poppitz2011,PhysRevLett.115.041601} pointed out that the complex saddle points related to the bions are important to provide the right vacuum energy. 
Even for theories with real couplings, complexifying the path integrals is always necessary \cite{PhysRevLett.116.011601}. 

In loop quantum gravity \cite{book,book1,han2007fundamental}, a recent result \cite{han2021complex} has shown that the key to solving the long-existing flatness problem \cite{Hellmann:2012kz,Engle:2020ffj,Bonzom:2009hw,Han:2013hna,Gozzini:2021kbt} is to find the complex saddle points of the analytically continued EPRL spin foam action \cite{EPRL,rovelli2014covariant}. 
These complex saddle points dominate the whole path integral when curvature exists; they are also categorized and endowed with geometrical interpretations \cite{Han:2021rjo}. 
Another recent result \cite{PhysRevD.103.084026} has used the Lefschetz thimbles attached to the complex saddle points as the integral cycles to numerically compute the correlation functions in the spin foam model. 
As such, studying the properties of the complex saddle points is necessary in a wide range of physical theories. 


For a complicated action, e.g., the spin foam action, solving the saddle point equation analytically can hardly be possible. 
This paper thus develops a numerical saddle-point finder that possesses the following characters\footnote{The saddle point method only applies to non-degenerate saddle points where the determinant of the hessian is not zero. This paper considers non-degenerate saddle points only.}:
\begin{itemize}
    \item working for complex valued action,
    \item being able to find saddle points without analytically solving the saddle point equation,
    \item being able to estimate the contribution of each saddle point to the partition function.
\end{itemize}
To work for complexified path integrals, our saddle-point finder combines the generalized thimble method (GTM) \cite{Alexandru:2020wrj} and a perturbative saddle-point finder (PSPF). 
The GTM uses the Lefschetz thimbles as integral cycles in a path integral to suppress the oscillation of the integrand in the complex-valued action.
On Lefschetz thimbles, the GTM samples points by the distribution $e^{S_{eff}}$, where the effective action $S_{eff}$ sums the real part of the action $S$ and the logarithm of the real part of the Jacobian caused by the deformation of the integral cycle.
Sampled points with significant statistical weights should be close to and thus can roughly locate the saddle points of $S$. 
At a sampled point, our PSPF finds where the local minimal value of $|\partial_\mu S|$ can be taken. 
Therefore, PSPF pins the saddle points around the sampled points. 
After finding the saddle points, one can compute the real part of the action at each of these saddle points to estimate its contribution to the whole partition function. 

In this paper, we showcase our saddle-point finder by two examples in the EPRL spin foam model: the single vertex case and the case of triangulation $\Delta_3$. 
In contrast to the method in \cite{han2021complex}, which only applies to the small deficit angle case, our finder can find multiple complex saddle points in the large deficit angle case. Furthermore, we find that in the large deficit angle case, multiple complex saddle points contribute to the spin foam amplitude and list these saddle points by their contributions to the partition function.

The paper is organised as follows. Section 2 reviews the GTM.
Section 3 introduces our saddle-point finder. 
Section 4 reviews the analytically continued spin foam model.
Sections 5 and 6 apply our saddle-point finder to the single-vertex EPRL spin foam and $\Delta_3$-triangulated EPRL spin foam.
Finally, Section 7 concludes the paper.

\section{Lefschetz thimble}
A Lefschetz thimble is a multi-dimensional generalization of the stationary phase contour of  a single-variable complex function. 
\cite{Witten:2010cx,Witten:2010zr} use the thimble method to define a new type of partition functions as integrals over thimbles instead of over $\mathbb{R}^N$.
Thimble method is also used in the asymptotic analysis related to the resurgent trans-series \cite{ANICETO20191}.
Numerically, the thimble method is used to compute observables when the action is complex valued \cite[etc]{Alexandru:2020wrj,Cristoforetti:2013qaa,Alexandru:2017czx,Alexandru:2015xva,Alexandru:2018ngw,Alexandru:2016san,Alexandru:2016ejd}.
For us, the thimble method can help roughly estimate the positions of saddle points.

One of the most important properties of the thimbles is that the imaginary part of the action is a constant on each thimble.
Therefore, the path integral along thimbles are not oscillatory.  
Assume a complex valued action $S$ of a lattice model.
One of the most useful integrals in the path integral formulation reads
\begin{equation}\label{eq:genform}
    F =\int \dd^N x O(\mathbf{x}) \e^{-S(\mathbf{x})}.
\end{equation}
When $O$ is $1$, $F$ is the partition function. 
To apply the thimble method, one has to first analytically continue $O(\mathbf{x})$ and $S(\mathbf{x})$ to be holomorphic functions $\hat{O}(\mathbf{z})$ and $\hat{S}(\mathbf{z})$, such that  
\be
    F =\int_{\mathbb{R}^N} \dd^N z \hat{O}(\mathbf{z}) \e^{-\hat{S}(\mathbf{z})}. 
\ee
The Picard-Lefschetz theory shows that the integral can be equivalently decomposed into a linear combination of integrals over $N$-dimensional integral cycles $\mathcal{J}_\sigma,\ \sigma=1\cdots N$:
\be
F=\sum_{\sigma} n_\sigma \int_{\mathcal{J}_\sigma} \dd^N z \hat{O}(\mathbf{z}) \mathrm{e}^{-\hat{S}(\mathbf{z})},
\ee
where $\mathcal{J}_\sigma$ labels the Lefschetz thimbles, and $n_\sigma$ labels the weight of each thimble.
Each thimble $\mathcal{J}_\sigma$ is defined as a union of the steepest decent (SD) paths meeting two conditions:
\begin{enumerate}
    \item Each path $z(t)$ is a solution to the SD equation 
  \begin{equation}\label{eq:SDeq}
      \frac{\dd z^a }{ \dd t}=-\frac{\partial\overline{\hat{S}(\mathbf{z})}}{\partial\overline{z^a}},
  \end{equation}
  where $z^a$ are the coordinates of the point $\mathbf{z}(t)$.
  \item On each path, $\mathbf{z}(t)$ goes to a saddle point $\mathbf{p}_\sigma$ when $t\to\infty$.
\end{enumerate}

Because 
\be
  \frac{\dd \hat{S}}{\dd t}=\frac{\partial \hat{S}}{\partial z^a} \frac{\dd z^a}{\dd t}=-\left\vert\frac{\partial \hat{S} }{\partial z^a} \right\vert^2,
\ee
$\mathrm{Re}(\hat{S})$ monotonically decreases along each SD path and approaches its minimum at the saddle point; $\mathrm{Im}(\hat{S})$ is conserved along each SD path. 
Therefore, on each thimble, the phase of each integrand becomes a constant, and 
\be\label{eq:integralonThimble}
  \int_{\mathcal{J}_\sigma} \dd^N z \hat{O}(\mathbf{z}) \mathrm{e}^{-\hat{S}(\mathbf{z})}
  =\mathrm{e}^{-\ii\, {\rm Im}{(\hat{S}(p_{\sigma}))}}\int_{\mathcal{J}_\sigma} \dd^N z \hat{O}(\mathbf{z}) \mathrm{e}^{-{\rm Re}{(\hat{S}(\mathbf{z}))}},
\ee
where the factor $\mathrm{e}^{-{\rm Re}{(\hat{S}(\mathbf{z}))}}$ is non-oscillatory now.
As a result, the oscillatory integral $F$ is equivalent to a combination of certain non-oscillatory integrals.

Ideally, the thimble method can be used to compute observables in the cases when only one thimble dominate to the whole path integral. An observable $\langle O \rangle$ reads
\be
 \langle O \rangle 
 = \frac{
    \int \dd^N x O(\mathbf{x}) \e^{-S(\mathbf{x})}
 }{
    \int \dd^N x \e^{-S(\mathbf{x})}
 }.
\ee
By the thimble method, 
\be
 \langle O \rangle 
 = \frac{
    \int \dd^N z \hat{O}(\mathbf{z}) \e^{-\hat{S}(\mathbf{z})}
 }{
    \int \dd^N z \e^{-\hat{S}(\mathbf{z})}
 }
 = \frac{
    \sum_{\sigma} n_\sigma \mathrm{e}^{-\ii\, {\rm Im}{(\hat{S}(p_{\sigma}))}} \int_{\mathcal{J}_\sigma} \dd^N z \hat{O}(\mathbf{z}) \e^{-{\rm{Re}}{(\hat{S}(\mathbf{z}))}}
 }{
    \sum_{\sigma} n_\sigma \mathrm{e}^{-\ii\, {\rm Im}{(\hat{S}(p_{\sigma}))}} \int_{\mathcal{J}_\sigma} \dd^N z \e^{-{\rm{Re}}{(\hat{S}(\mathbf{z}))}}
 }.
\ee
Assuming the thimble $\mathcal{J}_{\sigma'}$ governs the whole integral, $\langle O \rangle$ becomes
\be\label{eq:obs1}
\langle O \rangle
= \frac{
     \int_{\mathcal{J}_{\sigma'}} \dd^N z \hat{O}(\mathbf{z}) \e^{-{\rm{Re}}{(\hat{S}(\mathbf{z}))}}
 }{
     \int_{\mathcal{J}_{\sigma'}}  \dd^N z \e^{-{\rm{Re}}{(\hat{S}(\mathbf{z}))}}
 },
\ee
whose nominator and denominator are both non-oscillatory integrals.
In this case, the Lefschetz thimble method turns an oscillatory path integral into a statistical-mechanics problem. 
In fact, $\int_{\mathcal{J}_{\sigma'}} \dd^N z \e^{-{\rm{Re}}{(\hat{S}(\mathbf{z}))}}$ shown in \eqref{eq:obs1} is a partition function denoted as $Z$, where $\e^{-{\rm{Re}}{(\hat{S}(\mathbf{z}))}}$ can be considered as a Boltzmann factor. 
Such a statistical-mechanical system can be simulated by the Markov chain Monte Carlo (MCMC) method that samples points on the thimble $\mathcal{J}_{\sigma'}$ by the distribution $\e^{-{\rm{Re}}{(\hat{S}(\mathbf{z}))}}/Z$, and $\langle O \rangle$ is the mean value of $\hat{O}(\mathbf{z})$ among these sampled points. 
Note that $\mathrm{Re}(S)$ decreases along the SD paths, so the point possessing the largest $\e^{-{\rm{Re}}{(\hat{S}(\mathbf{z}))}}$ on $\mathcal{J}_{\sigma'}$ should be the saddle point $\mathbf{p}_{\sigma'}$. 
Thus, most sampled points should cluster around the saddle point.  

This ideal way to compute $\langle O \rangle$ is practically hard to be realized because 
\begin{itemize}
    \item in many cases, multiple thimbles contribute non-negligibly to  $\langle O \rangle$,
    \item and it is impossible to find the thimbles by solving the SD equation \eqref{eq:SDeq} with $t\to\infty$ in computers. 
\end{itemize}
Therefore, GTM has been developed to do the computation.
Instead of using the SD equation, GTM uses the steepest ascend (SA) equation  
\be\label{eq:SAequation}
    \frac{\dd z^a }{ \dd t}=\frac{\partial\overline{\hat{S}(\mathbf{z})}}{\partial\overline{z^a}}
\ee
to approach the thimbles. 
Let $\mathbf{z}(t)$ be a solution to \eqref{eq:SAequation} and $\mathbf{x}=\mathbf{z}(0)$.
Define $\mathcal{F}_T(\mathbf{x}):=\mathbf{z}(T)$.
An $N$-dimensional manifold $\mathcal{M}_T$ can be defined as $\{\mathcal{F}_T(\mathbf{x})| \mathbf{x}\in\mathbb{R}^N \}$. 
By Cauchy's theorem, 
\be
    F =\int_{\mathbb{R}^N} \dd^N z \hat{O}(\mathbf{z}) \e^{-\hat{S}(\mathbf{z})}=\int_{\mathcal{M}_T} \dd^N z \hat{O}(z) \e^{-\hat{S}(\mathbf{z})},
\ee
where the deformation from $\mathcal{M}_0=\mathbb{R}^N$ to $\mathcal{M}_T$ is continues.
According to \cite{Alexandru:2020wrj}, in the limit $T\to\infty$,  $\mathcal{M}_{T\to\infty}=\sum_{\sigma} n_\sigma \mathcal{J}_\sigma$. Therefore, for $T$ large enough,
\be\label{eq:LTL2}
    F
    =\sum_{\sigma} n_\sigma \int_{\mathcal{J}_\sigma} \dd^N z \hat{O}(\mathbf{z}) \mathrm{e}^{-\hat{S}(\mathbf{z})}
    \sim\int_{\mathcal{M}_T} \dd^N z \hat{O}(\mathbf{z}) \e^{-\hat{S}(\mathbf{z})}.
\ee

\begin{figure}[htbp]
    \centering
    \includegraphics[width=0.75\linewidth]{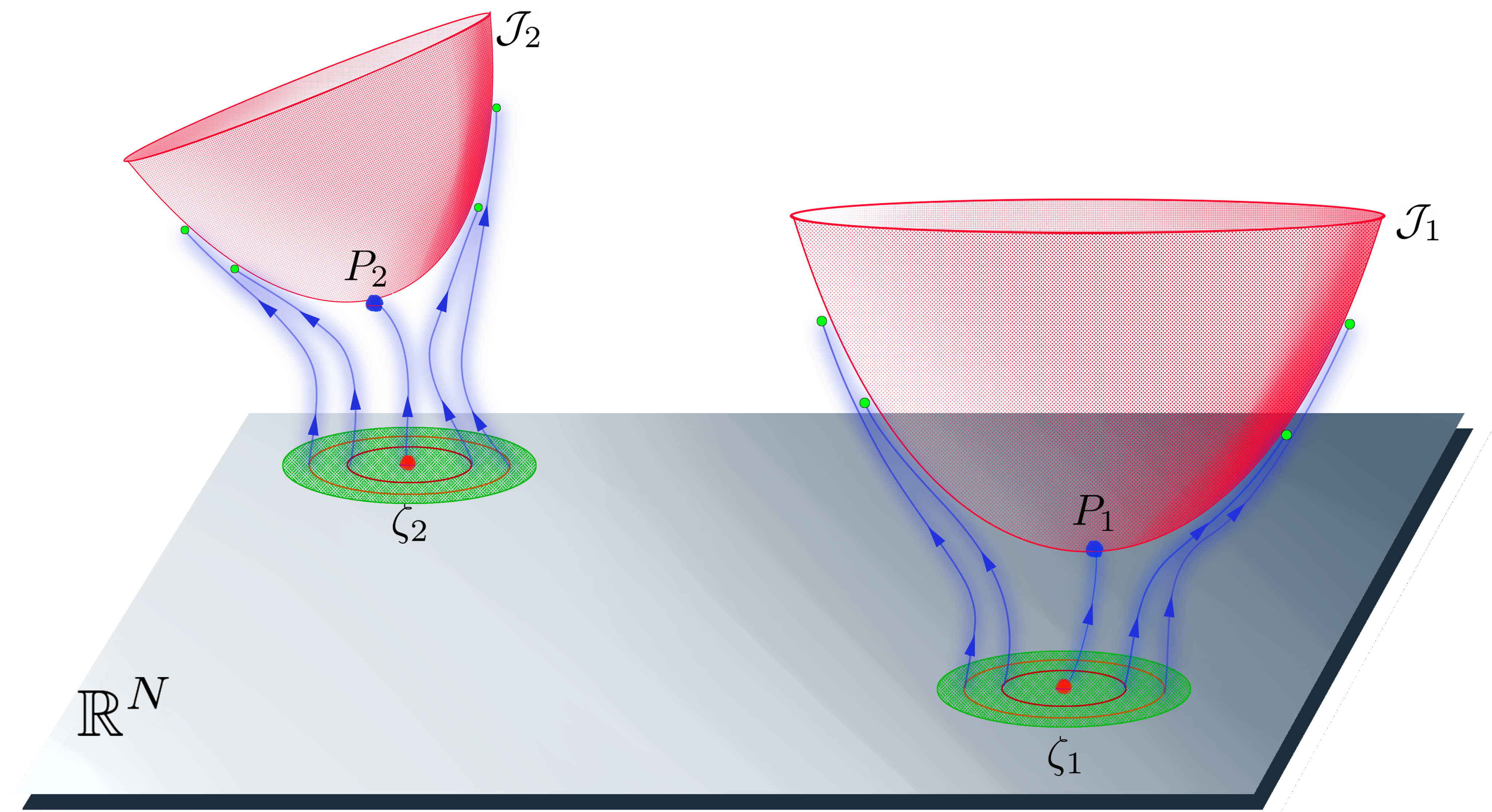}
    \caption{The grey plate indicates $\mathbb{R}^N$, The red manifolds $\mathcal{J}_1$ and $\mathcal{J}_2$ are two Lefschetz thimbles. The blue arrows indicate the SA flow. By SA flow, $\zeta_1$ and $\zeta_2$ are mapped to the saddle points $\zeta_1$  and $\zeta_2$, and the points in the green disks around $\zeta_1$ and $\zeta_2$ are mapped to the points close to the corresponding thimbles.}
    \label{fig:thimble}
\end{figure}

For some scattered points $\zeta\in\mathbb{R}^N$, $\mathcal{F}_T(\zeta)$ approach the saddle points $\mathbf{P}_\sigma$ of the thimbles with non-zero $n_\sigma$; for the points $\mathbf{x}\in\mathbb{R}^N$ close to $\zeta$, the set of $\mathcal{F}_T(\mathbf{x})$ forms an $N$-dimensional manifold approaching to the combination of $\mathcal{J}_\sigma$ (Fig. \ref{fig:thimble}).
To compute $\int_{\mathcal{M}_T} \dd^N z \hat{O}(\mathbf{z}) \e^{-\hat{S}(\mathbf{z})}$, we label each point $\mathcal{F}_T(\mathbf{x})\in\mathcal{M}_T$ by its initial point $x\in\mathbb{R}^N$ and transform $\int_{\mathcal{M}_T}$ back to $\int_{\mathbb{R}^N}$. 
 
Consider $\mathbb{R}^N=\mathcal{M}_0$, when $T=0$, $\partial \mathcal{F}_0(\mathbf{x})^k/\partial x^i=\delta_i^k$ defines the coordinate transformation from $\mathbb{R}^N$ to $\mathcal{M}_0$, and $\det{\delta}=1$ is the Jacobian for this coordinate transformation.
When $T\neq0$, the evolution of $\partial \mathcal{F}_t(\mathbf{x})^k/\partial x^i$ along an SA path is governed by 
\be\label{eq:jacobian1}
    \frac{\dd (\partial \mathcal{F}_t(\mathbf{x})^k/\partial x^i)}{\dd t}=\sum_{l=1}^n\overline{\frac{\partial^2 \hat{S}(\mathcal{F}_t(\mathbf{x}))}{\partial \mathcal{F}_t(\mathbf{x})^k \partial \mathcal{F}_t(\mathbf{x})^l }} (\overline{\partial \mathcal{F}_t(\mathbf{x})^l/\partial x^i}).
\ee
With the initial condition $\partial \mathcal{F}_0(\mathbf{x})^k/\partial x^i=\delta^k_i$, \eqref{eq:jacobian1} has the solution $\partial \mathcal{F}_T(\mathbf{x})^k/\partial x^i$, which describes the coordinate transformation from $\mathcal{F}_T(\zeta)$ to $\zeta$, with the Jacobian $J_T(\mathbf{x})=\det{(\partial \mathcal{F}_T(\mathbf{x})^k/\partial x^i)}$. As such, \eqref{eq:LTL2} becomes
\be
\label{eq:LTL}
    F
    \sim\int_{\mathcal{M}_T} \dd^N z \hat{O}(\mathbf{z}) \e^{-\hat{S}(\mathbf{z})}
    =\int_{\mathbb{R}^N} \dd^N x J_T(\mathbf{x}) \hat{O}(\mathcal{F}_T(\mathbf{x})) \e^{-\hat{S}(\mathcal{F}_T(\mathbf{x}))}.
\ee
Let ${S_T}_{eff}={\rm Re}(\hat{S}) - \log(\det(J_T))$ be the purely real effective action and ${\theta_T}_{res}=\arg(\det(J_T))-{\rm Im}(\hat{S})$ be the residual phase, \eqref{eq:LTL} becomes
\be\label{eq:LTL1}
     F
     \sim \int_{\mathbb{R}^N} \dd^N x  \hat{O}(\mathcal{F}_T(\mathbf{x})) \e^{\ii {\theta_T}_{res}} \e^{-{S_T}_{eff}(\mathcal{F}_T(\mathbf{x}))},
\ee
where $\e^{-{S_T}_{eff}(\mathcal{F}_T(\mathbf{x}))}$ can be considered as the Boltzmann factor of a sampling process on $\mathbb{R}^N$.
The observable \eqref{eq:obs1} can be computed by the re-weighted method \cite{Alexandru:2020wrj}:
\be\label{eq:rew}
\begin{split}
\langle O \rangle
\sim&\frac{
\int_{\mathbb{R}^N} \dd^N x  \hat{O}(\mathcal{F}_T(\mathbf{x})) \e^{\ii {\theta_T}_{res}} \e^{-{S_T}_{eff}(\mathcal{F}_T(\mathbf{x}))}
}{
\int_{\mathbb{R}^N} \dd^N x   \e^{\ii {\theta_T}_{res}} \e^{-{S_T}_{eff}(\mathcal{F}_T(\mathbf{x}))}
}\\
=&\frac{
\int_{\mathbb{R}^N} \dd^N x  \hat{O}(\mathcal{F}_T(\mathbf{x})) \e^{\ii {\theta_T}_{res}} \e^{-{S_T}_{eff}(\mathcal{F}_T(\mathbf{x}))}
}{
\int_{\mathbb{R}^N} \dd^N x    \e^{-{S_T}_{eff}(\mathcal{F}_T(\mathbf{x}))}
}
\frac{
\int_{\mathbb{R}^N} \dd^N x   \e^{-{S_T}_{eff}(\mathcal{F}_T(\mathbf{x}))}
}{
\int_{\mathbb{R}^N} \dd^N x   \e^{\ii {\theta_T}_{res}} \e^{-{S_T}_{eff}(\mathcal{F}_T(\mathbf{x}))}
}\\
=&\frac{
\langle \hat{O} \e^{\ii {\theta_T}_{res}} \rangle_{Teff}
}{
\langle \e^{\ii {\theta_T}_{res}} \rangle_{Teff}
}.
\end{split}
\ee
$\langle f\rangle_{Teff}$ is the mean value of any given $f$ among the sampled points.

Although the integrands in \eqref{eq:rew} are still oscillatory, the fluctuation is much smaller in $\mathcal{M}_T$ than in $\mathbb{R}^N$ for large $T$. 
In $\mathcal{M}_T$, the points with significant distribution come from small isolated regions around the saddle points.
In each such small region, $\e^{\ii {\theta_T}_{res}}$ oscillates mildly. 
Outside these small regions, $\e^{\ii {\theta_T}_{res}}$ oscillates severely, but the points here contribute little to the whole integral.
As a result, the larger $T$ is, the smaller the contributing regions are and the less oscillating the integrands are.
This property ensures that with properly chosen $T$, most the sampled points in the GTM are around the saddle points, and our saddle-point finder uses this fact.

Besides, the choice of $T$ is important in the GTM. 
On the one hand, large $T$ can suppress the oscillation of the integrands. 
On the other hand, the larger the $T$, the more isolated the contributing regions. Isolated regions are a landscape that is hard to be sampled by samplers like MCMC or slice sampling. 
For a multi-modal distribution with multiple contributing regions, the sampler depending on local movements may be trapped in one of the regions.
To resolve this issue, the worldvolume-tempered Lefschetz thimble method (WV-TLTM) has been developed \cite{fukuma2020worldvolume}.
By Cauchy's theorem, the value of $\langle O \rangle$ is independent of the choice of $T$:
\be
\langle O \rangle
\sim\frac{
\langle \hat{O} \e^{\ii {\theta_{T_1}}_{res}} \rangle_{{T_1}eff}
}{
\langle \e^{\ii {\theta_{T_1}}_{res}} \rangle_{{T_1}eff}
}
=\frac{
\langle \hat{O} \e^{\ii {\theta_{T_2}}_{res}} \rangle_{{T_2}eff}
}{
\langle \e^{\ii {\theta_{T_2}}_{res}} \rangle_{{T_2}eff}
}, \, T_1\neq T_2.
\ee
Therefore, $\langle O \rangle$ can be computed by considering the contributions of different $T$, i.e.,
\be
\begin{split}
\langle O \rangle
\sim&\frac{
\int_{T_0}^{T_1} \dd T \e^{-W(T)}\int_{\mathbb{R}^N} \dd^N x  \hat{O}(\mathcal{F}_T(\mathbf{x})) \e^{\ii {\theta_T}_{res}} \e^{-{S_T}_{eff}(\mathcal{F}_T(\mathbf{x}))}
}{
\int_{T_0}^{T_1} \dd T \e^{-W(T)} \int_{\mathbb{R}^N} \dd^N x   \e^{\ii {\theta_T}_{res}} \e^{-{S_T}_{eff}(\mathcal{F}_T(\mathbf{x}))}
},\\
\end{split}
\ee
where $W(T)$ is an arbitrary function. 
In this computation, the sampling is performed on the worldvolume defined as
\[
\mathcal{R}=\bigcup_{i=T_0}^{T_1}\mathcal{M}_T.
\]
In an $\mathcal{M}_T$ with small $T$, the contributing regions are so large that they will contact with each other, and the sampler may use this $\mathcal{M}_T$ as a bridge between the isolated regions in large $T$ slices. 
Therefore, by considering the interval between a small $T$ and a large $T$, WV-TLTM can sample over all the regions containing saddle points. 


\section{Saddle-Point Finder}
In our finder, saddle points are found by a two-level searching procedure. 
On the first level, the GTM serves as the coarse finder to roughly locate the saddle points. 
On the second level, the PSPF is deployed to pin the saddle points. 
This section introduces the coarse finder first and then the pinpoint finder.
\subsection{The coarse finder}
The GTM can sample the points around saddle points. 
Specifically, we use the ensemble slice sampling method \cite{Karamanis2021} as the sampler and WV-TLTM to combine the contributions of the different evolution time $T$.
The finder consists of the following steps:
\begin{enumerate}
    \item Choose $A$ points $\{ \mathbf{x}_i, i=1\cdots A \}$. If the action is a function depending on $N$ complex variables, $A>2N$ is suggested.
    \item Using $\{ \mathbf{x}_i, i=1\cdots A \}$ as initial points of the SA flow, find the maximal time $T_1$, till which the differential equation solver can evolve all these points. Pick a time $T_0<T_1$ and use $(T_0,T_1)$ as the time interval in WV-TLTM.
    \item Apply the ensemble slice sampling method (Algorithm \ref{alg:GTM1}) to sample on the worldvolume by the distribution density $\e^{-{S_T}_{eff}(\mathcal{F}_T(\mathbf{x}))}$.
    \item Sort the sampled points $x$ by their effective action. Take the first $P$ points as the output of the finder. Here, $P$ is a parameter of the finder, and it needs to be tuned to achieve the best performance.
\end{enumerate}

In the second step, any ODE solver cannot evolve the SA flow for infinitely long.
The right hand side of the \eqref{eq:SAequation} becomes larger and larger when the flow is leaving the saddle point. 
The ODE numerical solvers, e.g., Runge-Kutta, Rosenbrock, etc, use difference equations to approximate the differential equations. 
The error of this approximation is proportional to the norm of the right hand side of the differential equations.
Therefore, the error increases with the evolution time, and the maximal time $T_1$ is the largest evolution time, such that the error is under the given tolerance.

The ensemble slice sampling (ESS) used in the third step is a powerful MCMC sampler that applies to complicated cases. 
As a type of slice sampling \cite{10.1214/aos/1056562461}, the basic idea of ESS is that sampling from a distribution $p(x)$ whose density is proportional to $f(x)$ is equivalent to uniformly sampling from the region below the curve of $f(x)$. 
In many cases \cite{Karamanis2021}, ESS performs better than those random-walking based MCMC sampler for multimodal distribution, and we take this advantage of ESS to sample on the $\mathcal{M}_T$.
The ESS defines an ensemble $\{ \mathbf{x}_1,\cdots\mathbf{x}_A\}$ of parallel chains and generates moves by the positions of the current head of the chains $\{ \mathbf{x}_1^{(t)},\cdots\mathbf{x}_A^{(t)}\}$. 
In each ESS iteration, we first apply the differential move scheme to generate the direction vector for each chain $\mathbf{x}_k$. 
This scheme comprises two steps:
\begin{enumerate}
    \item From the complementary ensemble $S_{[k]}=\{\mathbf{x}_n,\ \forall n \neq k\}$, draw two chains $\mathbf{x}_l$ and $\mathbf{x}_m$ uniformly and without replacement.
    \item Compute the direction vector $\vec{\eta}_k$ by $\vec{\eta}_k=\mu(\mathbf{x}_l-\mathbf{x}_m)$.
\end{enumerate}
The parameter $\mu$ can be automatically tuned by the method in \cite{Karamanis2021}.
Then, we apply $\vec{\eta}_k$ in Algorithm \ref{alg:GTM1} to generate the moves for this ESS iteration.
\begin{figure}[h]
  \begin{algorithm}[H]
    \caption{Ensemble slice sampling}\label{alg:GTM1}
    \begin{algorithmic}[1]
     \STATE{Given $t$, $f$, $S$:}
    \STATE{Initialise $N_{e}^{(t)} = 0$ and $N_{c}^{(t)} = 0$}
    \FOR{$k=1, ..., A$}
        \STATE{Get direction vector $\vec{\eta}_{k}$}
        \STATE{Sample $Y \sim \text{Uniform}(0,f(\mathbf{x}_{k}^{(t)}))$}
        \STATE{Sample $U \sim \text{Uniform}(0,1)$}
        \STATE{Set $L \leftarrow - U$, and $R \leftarrow L + 1$}
        \WHILE{$Y < f(\mathbf{x}_{k}^{(t)}+L\vec{\eta}_{k})$}
            \STATE{$L \leftarrow L - 1$}
            \STATE{$N_{e}^{(t)} \leftarrow N_{e}^{(t)} + 1$}
        \ENDWHILE
        \WHILE{$Y < f(\mathbf{x}_{k}^{(t)}+R\vec{\eta}_{k})$}
            \STATE{$R \leftarrow R + 1$}
            \STATE{$N_{e}^{(t)} \leftarrow N_{e}^{(t)} + 1$}
        \ENDWHILE
        \WHILE{True}
            \STATE{Sample $X' \sim \text{Uniform}(L,R)$}
            \STATE{Set $Y' \leftarrow f(X'\vec{\eta}_{k} + \mathbf{x}_{k}^{(t)})$}
            \IF{$Y<Y'$}
                \STATE{\bf{break}}
            \ENDIF
            \IF{$X'<0$}
                \STATE{$L \leftarrow X'$}
                \STATE{$N_{c}^{(t)} \leftarrow N_{c}^{(t)} + 1$}
            \ELSE
                \STATE{$R \leftarrow X'$}
                \STATE{$N_{c}^{(t)} \leftarrow N_{c}^{(t)} + 1$}
            \ENDIF
        \ENDWHILE
        \STATE{Set $\mathbf{x}_{k}^{(t+1)} \leftarrow X' \eta_{k} + \mathbf{x}_{k}^{(t)}$}
    \ENDFOR
      \end{algorithmic}
  \end{algorithm}
  \end{figure}
The whole ESS sampling process consists of multiple ESS iterations.

In our work, the distribution $f(\mathbf{x})$ is chosen to be $\e^{-{S_T}_{eff}(\mathcal{F}_T(\mathbf{x}))}$, and the space for sampling is $\mathbb{R}^N$. 
We remark that although theoretically the ergodicity of WV-TLTM is proven, the efficiency of the sampling procedure can be very low for large $N$. 
We can improve the efficiency of the finder by the following pre-treatments:
\begin{itemize}
    \item Find a compact region of interest as the working place of the finder.
    \item Find the points with small value of $|\partial_\mu f(\mathbf{x})|$ within the compact region by physical facts or by optimization algorithm, e.g., annealing algorithm, genetic algorithm, particle swarm algorithm, etc.
\end{itemize}

\subsection{The pinpoint finder}
The coarse finder feeds multiple points around the saddle points to the pinpoint finder that applies the PSPF to locate the saddle points.
The PSPF is based on that there always exists a point $\tilde{x}$ such that $|\partial_\mu f (\tilde{x})| \le |\partial_\mu f (x)|$ for any $x\in\mathbb{C}^N$ and a locally smooth function $f(x)$ with $\det{\frac{\partial^2 f(x)}{\partial x^\mu \partial x^\nu}}\neq 0$.
Let $\vec{\epsilon}=-(\frac{\partial^2 f(x)}{\partial x^\mu \partial x^\nu})^{-1}\partial_\nu f(x)$, $\partial_{\mu}f(x+\alpha\epsilon)$ expands as
\be
\begin{split}
    \partial_\mu f(x+\alpha\epsilon)&=\partial_\mu f(x)+\alpha \frac{\partial^2 f(x)}{\partial x^\mu\partial x^\nu} \epsilon^\nu +O(\alpha^2\epsilon^2)\\
    &=\partial_\mu f(x)-\alpha  \frac{\partial^2 f(x)}{\partial x^\mu\partial x^\nu} (\frac{\partial^2 f(x)}{\partial x^\nu \partial x^\lambda})^{-1}\partial_\lambda f(x) +O(\alpha^2\epsilon^2)\\
    &=(1-\alpha) \partial_\mu f(x)+O(\alpha^2\epsilon^2).
\end{split}
\ee
Hence, for a positive but sufficiently small $\alpha$, $|\partial_\mu f(x+\alpha\epsilon)|<|\partial_\mu f(x)|$, then we can use $x+\alpha\epsilon$ as the $\tilde{x}$.
Recursively taking the output $\tilde{x}$ as the input $x$, one can find the location of the nearest local minimal value of $|\partial_\mu f|$ where $\det{\frac{\partial^2 f(x)}{\partial x^\mu \partial x^\nu}}\neq 0$.
Algorithm \ref{alg:PTF} with three parameters ($N$, $toa$, $tol$) summarizes the PSPF method.
\begin{figure}[h]
  \begin{algorithm}[H]
    \caption{Perturbative Finder}\label{alg:PTF}
    \begin{algorithmic}[1]
        \STATE Given parameters $N$, $toa$, and $tol$ and function $f$:
        \STATE Initialise $k=0$
        \WHILE{$k<N$}
            \STATE  $\epsilon \leftarrow -(f''(x_0))^{-1}\cdot f'(x_0)$
            \STATE  $\alpha\leftarrow 0$
            \WHILE{ $\alpha<toa$ }
            \STATE $C \leftarrow |f'(x_0)|-|f'(x_0+ 10^{-\alpha}\times\epsilon)|$ 
            \IF{C > 0}
                \STATE{\bf{break}}
            \ENDIF
            \STATE $\alpha\leftarrow \alpha+1$
            \ENDWHILE
            \IF{$C<tol$}
                \STATE{\bf{break}}
            \ENDIF
            \STATE {$x_0\leftarrow x_0+ 10^{-\alpha}\times\epsilon$}
            \STATE $k\leftarrow k+1$
        \ENDWHILE
    \end{algorithmic}
   \end{algorithm}
 \end{figure}
 
The parameter $N$ defines the upper limit of the number of iterations;
$toa$ and $tol$ are the lower bounds of $\alpha$ and $|\alpha \epsilon|$; $toa$ indicates the accuracy of the algorithm. The algorithm terminates when the PSPF finds $||f(x_0)|-|f(\tilde{x})||< tol$. 
\begin{figure}[htpb]
    \centering
    \includegraphics[width=0.9\linewidth]{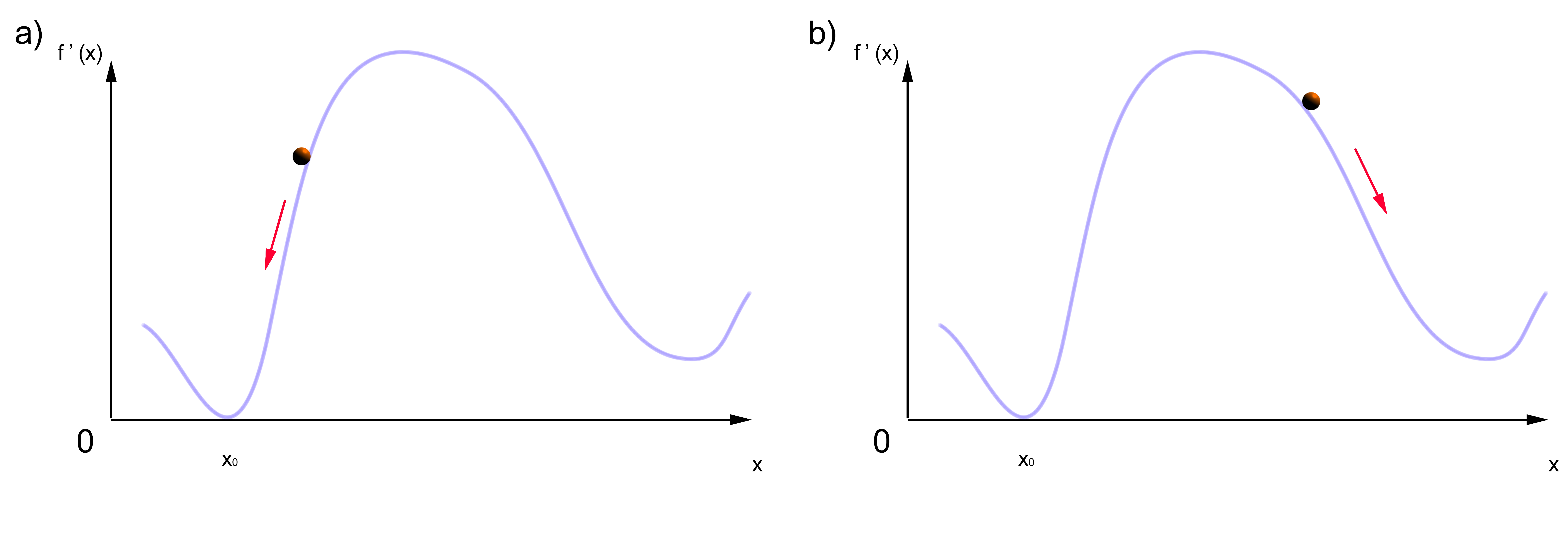}
    \caption{For both figures, the vertical axis corresponds to the $f'(x)$ and the horizontal axis corresponds to the $x$. For the a), the red point is close to the saddle point $x_0$, and PSPF can move the red points to $x_0$. For the b), there is a bump between the saddle point $x_0$ and the red point, and PSPF cannot move the red point to the saddle point.}
    \label{fig:pet}
\end{figure}
For a point far away from all the saddle points (Fig. \ref{fig:pet} (b)), the PSPF cannot find the saddle points. 
Nevertheless, when a point is close to one of the saddle points (Fig. \ref{fig:pet} (a)) the PSPF can find saddle points.
Consequently, pinpoint finder can locate the saddle points from most points fed by the coarse finder.

Finally, we note that our pinpoint finder can only work for the cases with $\det{\frac{\partial^2 f(x)}{\partial x^\mu \partial x^\nu}}\neq 0$. 
Therefore, the finder can only find non-degenerate saddle points. 
In fact, in the Lefschetz thimble method, thimbles attach only to non-degenerate saddle points, and degenerate saddle points do not contribute to the partition function.
\section{The analytically continued spin foam model}
Spin foam is a covariant formulation of loop quantum gravity \cite{EPRL,Freidel:2007py,Reisenberger:2000fy,Perez:2003vx,carlobook}. 
In this work, we use the EPRL spin foam model \cite{EPRL} as the proving ground of our saddle-point finder. Here, we review the action of the EPRL spin foam model, the analytic continuation of the EPRL action, and the classification of the complex saddle points of the analytically continued action.

The partition function of the spin foam model is often called the spin foam amplitude.
The spin foam amplitude depends on the boundary spin-network state. 
The general form of the spin foam amplitude on a simplicial complex $\mathcal{K}$ reads 
\be\label{eq:unif}
Z=\sum_{\vec{J}}\prod_{f} \mathbf{d}_{J_f} \int [\dd X] \e^{\sum_f J_fF_f[X,T]},
\ee
where $f$ labels the 2-faces in $\mathcal{K}$ colored by spins $J_f$, $\sum_{\vec{J}}$ means summing over all the possible ways of coloring $\mathcal{K}$ by spins, $X$ collects all the variables to be integrated, $T$ collects the parameters determined by the given boundary state, and $\sum_f J_fF_f[X,T]$ is the action.

\begin{figure}[!ht]
\centering\includegraphics[width=1\linewidth]{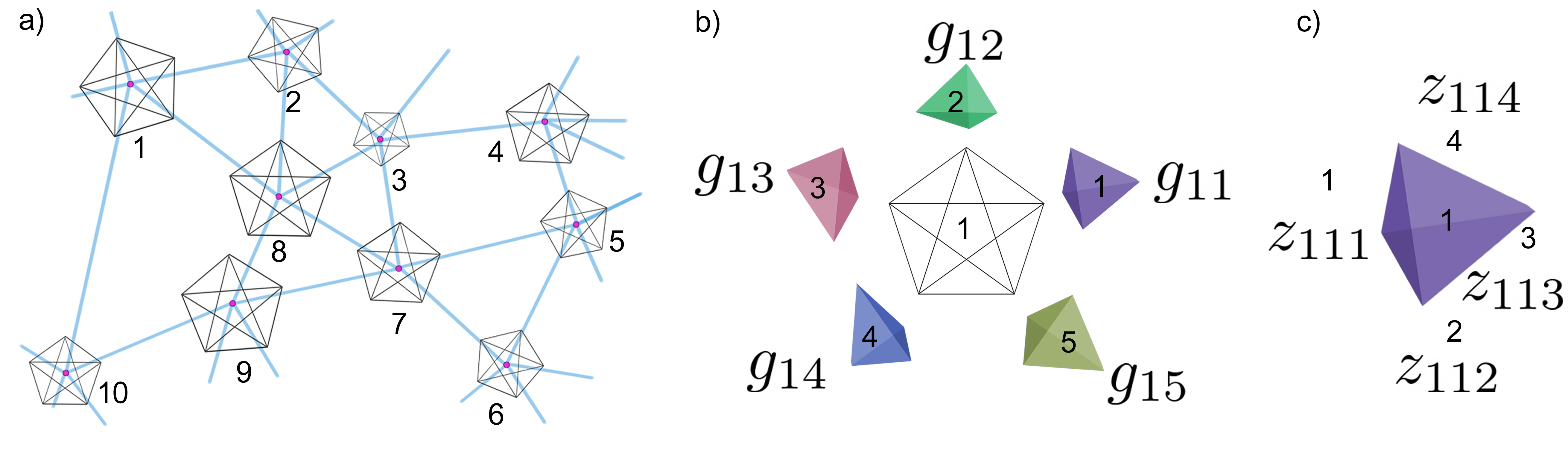}
\caption{a)In a simplical complex, the $4$-simplices are labelled by $v,\, v=1,\cdots, 10$. b) For $4$-simplex $1$, five tetrahedra are labelled by $e,\, e=1,\cdots, 5$. Each tetrahedron are assigned with a group variable $g_{ve}$, where $ve$ is $12,\,13,\,14,\,15$. c) For the tetrahedron $ve=11$, the faces are labelled by $111,\,112,\,113,\,114$, and each face is assigned by a spinor.  }\label{fig:complexz}
\end{figure}
In the Lorentzian EPRL model, 
\be\label{eq:setup1}
\left\{
\begin{aligned}
    \mathbf{d}_{J_f}&=2J_f+1;\\
    X&\equiv (g_{ve},z_{vf},\xi_{ef}^I);\\
    \dd X & \equiv \dd g_{ve} \dd \tilde{z}_{vf} \dd \xi_{ef}^I;\\
    T&\equiv (\xi_{ef}^B).\\
\end{aligned}
\right.
\ee
Here, $v$ denotes a $4$-simplex in $\mathcal{K}$, and each $3$-dimensional tetrahedron in $\partial v$ is denoted by $e$.
For each $v$, the group variables $g_{ve}\in\SLDC$ are assigned to tetrahedra; the spinor variables $z_{vf}\in\mathbb{CP}^1$ are assigned to the faces (see Fig.\ref{fig:complexz}). 
Both $\xi_{ef}^I$ and $\xi_{ef}^B$ are $\mathbb{C}^2$ spinors normalized by Hermitian inner product. 
Variables $\xi_{ef}^I$, which are assigned to the internal faces in $\mathcal{K}$, need to be integrated.
Parameters $\xi_{ef}^B$, which are assigned to the boundary faces, are fixed by the boundary states.
The $\SLDC$ Haar measure $\dd g_{ve}$ can be expressed as \cite{hanPI}
\be
\mathrm{d} g=\frac{\mathrm{d} \beta \mathrm{d} \beta^{*} \mathrm{~d} \gamma \mathrm{d} \gamma^{*} \mathrm{~d} \delta \mathrm{d} \delta^{*}}{|\delta|^{2}} \quad \forall g=\left(\begin{array}{cc}
\alpha & \beta \\
\gamma & \delta
\end{array}\right).
\ee
Let $Z_{vef}=g^\dagger_{ve}z_{vf}$ and $\langle\cdot,\cdot\rangle$ be the $\Su$ invariant inner product, $\forall z_{vf}=(z_0,z_1)$, 
\be
\begin{split}
\mathrm{d} \tilde{z}_{vf}
&=-\frac{
\mathrm{d} z_{vf}
}{
\left\langle Z_{vef}, Z_{vef}\right\rangle\left\langle Z_{ve'f}, Z_{ve'f}\right\rangle
}\\
&=-\frac{i}{2}\frac{
\left(z_{0} \dd z_{1}-z_{1} \dd z_{0}\right) \wedge\left(\bar{z}_{0} d \bar{z}_{1}-\bar{z}_{1} d \bar{z}_{0}\right)
}{
\left\langle Z_{vef}, Z_{vef}\right\rangle\left\langle Z_{ve'f}, Z_{ve'f}\right\rangle
}.\\
\end{split}
\ee
Here, $e,e' \in \partial v$ are two tetrahedra sharing the face $f$.
Let $\{v| f \subset v\}$ as the set of $4$-simplices containing the face $f$,
\be
F_{f}[X,T]=\sum_{\{v| f \subset v\}}\left(\ln \frac{\left\langle\xi_{e f}, Z_{v e f}\right\rangle^{2}\left\langle Z_{v e^{\prime} f}, \xi_{e^{\prime} f}\right\rangle^{2}}{\left\langle Z_{v e f}, Z_{v e f}\right\rangle\left\langle Z_{v e^{\prime} f}, Z_{v e^{\prime} f}\right\rangle}-\mathrm{i} \gamma \ln \frac{\left\langle Z_{v e f}, Z_{v e f}\right\rangle}{\left\langle Z_{v e^{\prime} f}, Z_{v e^{\prime} f}\right\rangle}\right),
\ee
where $\gamma$ is the Barbero-Immirzi parameter. 
Depending on $\mathcal{K}$, $\xi_{e f}$ can be either $\xi_{e f}^I$ or $\xi_{e f}^B$.
By the convention in [cites], some of the $\xi_{e f}$ in $F_f$ can be replaced by $J\xi_{e f}$ where $J\xi=(\bar{\xi}_2,-\bar{\xi}_1)$ for a spinor $\xi=(\xi_1,\xi_2)$.

The EPRL spin foam action has two types of gauge degrees of freedom---the continuous gauges and the discrete gauge \cite{hanPI}.
There are three continuous gauge degrees of freedom:
\begin{enumerate}
\item rescaling of $z_{vf}$:
\be
z_{vf}\mapsto \lambda z_{vf},\ \ \ \lambda\in\C;
\ee
\item $\SLDC$ gauge transformation at each $4$-simplex $v$:
\be\label{eq:gsldc}
g_{ve}\mapsto x^{-1}_vg_{ve},\ \ \ z_{vf}\mapsto x^\dagger_vz_{vf},\ \ \ x_v\in\SLDC;
\ee
\item $\Su$ gauge transformation on each internal tetrahedron $e$, i.e., the tetrahedron shared by two 4-simplices:
\be\label{eq:gsu}
g_{ve}\mapsto g_{ve}h_e^{-1},\ \ \ h_e\in\Su.
\ee
\end{enumerate}
The discrete gauge is flipping the sign of the group variables $g_{ve}\mapsto -g_{ve}$. 
The group variables take value of Lorentz group $\mathrm{SO}^+(1,3)$ rather than its double-cover $\SLDC$.

In our work, we parameterize the EPRL spin foam action after fixing the continuous gauges. 
By fixing the rescaling gauge of $z_{vf}$, each $z_{vf}$ can be parameterized by two real variables: 
\be\label{eq:spinorp}
z_{vf}=\left(1, x_{vf}+\ii y_{vf} \right).
\ee
By fixing the $\SLDC$ gauge in each 4-simplex, one can set one of the five $g_{ve}$ at each vertex $v$ as identity.
For any $\SLDC$ group element $g$, one can always decompose $g$ into $g'h$ where $h$ is an $\Su$ element and $g'$ is a triangular matrix. 
Thus, to fix the $\Su$ gauge in the internal tetrahedron $e$, one can parameterize one of two $\SLDC$ elements assigned to $e$ as 
\be\label{eq:sup}
\left(
\begin{matrix}
\lambda^{-1}&  x+\ii y\\
0           &  \lambda
\end{matrix}
\right),\ \ \lambda \in \mathbb{R}\setminus\{0\},\, x,y\in \mathbb{R}
\ee
and parameterize the other $\SLDC$ element as
\be\label{eq:slp}
\left(
\begin{matrix}
1+(x_1+\ii y_1)/\sqrt{2} &  (x_2+\ii y_2)/\sqrt{2}\\
(x_3+\ii y_3)/\sqrt{2}   &  \frac{1+(x_2+\ii y_2)(x_3+\ii y_3)/2}{1+(x_1+\ii y_1)/\sqrt{2}}\\
\end{matrix}
\right),\ \ x_1,y_1,x_2,y_2,x_3,y_3\in \mathbb{R}.
\ee
For each boundary tetrahedron, the assigned $\SLDC$ element is also parameterized as \eqref{eq:slp} too.

It is convenient to shift one of the saddle points to the origin $x=y=0$\footnote{Here, $x$ and $y$ stand for all real variables in \eqref{eq:slp}, \eqref{eq:sup}, and \eqref{eq:spinorp}.}.
Denoting $(1,{z_0}_{vf})$ and ${g_{0}}_{ve}$ as the saddle point value of $z_{vf}$ and $g_{ve}$, \eqref{eq:spinorp}, \eqref{eq:sup}, and \eqref{eq:slp} can be modified as
\be\label{eq:cepar}
\begin{split}
  z_{vf}&=\left(1, {z_0}_{vf}+x_{vf}+\ii y_{vf} \right),\\
  g_{ve}&={g_{0}}_{ve}\left(
\begin{matrix}
\lambda^{-1}&  x+\ii y\\
0           &  \lambda
\end{matrix}
\right),\\
g_{ve}&={g_{0}}_{ve}\left(
\begin{matrix}
1+(x_1+\ii y_1)/\sqrt{2} &  (x_2+\ii y_2)/\sqrt{2}\\
(x_3+\ii y_3)/\sqrt{2}   &  \frac{1+(x_2+\ii y_2)(x_3+\ii y_3)/2}{1+(x_1+\ii y_1)/\sqrt{2}}\\
\end{matrix}
\right).\\
\end{split}
\ee
With the parameterization defined by \eqref{eq:spinorp}, \eqref{eq:sup}, and \eqref{eq:slp}, the measure $\dd g_{ve}$ and $\dd z_{vf}$ become
\[
dg=\frac{1}{128\pi^4}\frac{dx_1dx_2dx_3dy_1dy_2dy_3}{|1+\frac{x_1+\ii y_1}{\sqrt{2}}|^2},
\]
\[
\dd z_{vf}=\dd x_{vf} \dd y_{vf}.
\]

The analytic continuation of $F_f$ can be realized by complexifing the group variables $g_{ve}$ and the spinor variables $z_{vf}$.
After this complexification, $g_{ve}\in\SLDC$ becomes ${\tilde{g}}_{ve}\in SO(4,\mathbb{C})$, and $g_{ve}^\dagger\in\SLDC$ becomes ${\tilde{g}'}_{ve}\in SO(4,\mathbb{C})$, which is independent of $\tilde{g}_{ve}$. 
Similarly, the spinor $z_{vf}\in\mathbb{CP}^1$ becomes $\tilde{z}_{vf}\in\mathbb{C}^2$, and $\bar{z}_{vf}$ becomes ${\tilde{z}'}_{vf}\in\mathbb{C}^2$, which is independent of $\tilde{z}_{vf}$. 
This analytic continuation complexifies all the real parameters appearing in \eqref{eq:spinorp}, \eqref{eq:sup}, and \eqref{eq:slp}.

The analytically continued EPRL spin foam action brings in three types of complex saddle points corresponding to the following three types of geometry
\begin{enumerate}
    \item {\bf{Non-degenerate simplicial geometry:}}
    Each vertex indicates a $4$-simplex. Each of the $10$ faces is represented by a bivector $B^f$. For each tetrahedron, the bivectors of the four faces satisfy the closure condition:
    \be
    \sum_{j,j\neq i} B^f_{ij}=0,
    \ee
    The volume $V_i$ of tetrahedron $i$ is non-zero. 
    Each tetrahedron has a $4$-dimensional normal vector $N^i$:
    \be
    N^i B^f_{ij}=0.
    \ee
    This condition is also known as the cross simplicial condition [cite].
    The $4$-dimensional normal vectors fulfill the $4$-dimensional closure condition:
    \be
    \sum_i V_i N_i = \sum_i U_i=0.
    \ee
    The volume of the $4$-simplex is non-zero:
    \be
    v_a=\frac{5!}{\sum_{ijkl}\epsilon_{ijkl}\det{[U_i,U_j,U_k,U_l]}}\neq0.
    \ee
    \item {\bf{Degenerate vector geometry:}}
    For a $v$ interpreted as vector geometry, there exist $10$ bivectors corresponding to $10$ faces, and they all belong to the same $3$-dimensional subspace. For each tetrahedron, the closure condition and cross simplicial condition hold; however, the $4$-dimensional normal vectors of the five tetrahedra are parallel to each other.
    Therefore, the volume of $v$ is ill-defined, rendering this type of geometry degenerate.
    \item {\bf{Lorentzian $SO(1,3)$ bivector geometry:}}
    On each $v$, this type of geometry also depends on $10$ faces represented by bivectors.
    These bivectors fulfill the closure condition but not the cross simplicial condition. 
    This indicates that those $10$ faces cannot form $5$ tetrahedra as required by the simplicial geometry.
\end{enumerate}
This classification depends crucially on the behavior of the $4$-dimensional normal vectors. 
At each saddle point, one can always try to reconstruct the $4$-dimensional normal vectors by $\tilde{g}_{ve}$, $\tilde{g}'_{ve}$, $\tilde{z}_{vf}$, and ${\tilde{z}'}_{vf}$. 
If the $4$-dimensional normal vectors at a saddle point cannot be reconstructed, the saddle point indicates an $SO(1,3)$ bivector geometry. 
If the reconstructed normal vectors at a saddle point are parallel to each other, the saddle point indicates a vector geometry, and if they make $\epsilon_{ijkl}\det{[U_i,U_j,U_k,U_l]}$ non-zero, the saddle point indicates a simplicial geometry.
\section{Application: The saddle points in the single $4$-simplex spin foam model}
\subsection{The action}
The first example of applying our saddle point finder is the single-vertex spin foam model. 
This model describes how five space-like quantum tetrahedra interact with each other.
In this model, we only have one $4$-simplex, so we can neglect the $v$ label in this section. 
The index $a$ labels the tetrahedra, and the index pair $ab$ labels the face shared by two tetrahedra $a$ and $b$.
The index $a$ runs from $1$ to $5$ because a $4$-simplex has five boundary tetrahedra.
All the faces in a $4$-simplex are boundary faces, whose geometric information is encoded in the parameters $\xi_{ab}$.
Following \cite{propagator,propagator1,propagator2,propagator3}, we use a coherent spin-network state as the boundary state, such that \eqref{eq:unif} takes the form
\be\label{eq:amp1}
Z=\sum_{\vec{J}}\psi_{ J_0,\zeta_0}\prod_{ab} \mathbf{d}_{J_{ab}} \int [\dd X] \e^{\sum_{a>b} J_{ab}F_{ab}[X,T]}.
\ee
In this section, we have
\be
\left\{
\begin{split}
    \mathbf{d}_{J_{ab}}&=2J_{ab}+1,\\
    X&\equiv (g_{a},z_{ab},J_{ab}),\\
    \dd X & \equiv (\dd g_{a}, \dd \tilde{z}_{ab}),\\
    T&\equiv (\xi_{ab},\zeta_{0}^{ab},{J_0}_{ab},\alpha^{(ab)(cd)}),\\
\end{split}
\right.
\ee
\be
\begin{split}
    \psi_{J_0,\zeta_0}=&\exp\left( -i \sum_{ab} \zeta_0^{ab}(J_{ab}-{J_{0}}_{ab})\right) \\
    \times&\exp\left(-\sum_{ab,cd}\alpha^{(ab)(cd)} \frac{J_{ab}-{J_{0}}_{ab}}{\sqrt{{J_{0}}_{ab}}}\frac{J_{cd}-{J_{0}}_{cd}}{\sqrt{{J_{0}}_{cd}}}\right),
\end{split}
\ee
and
\be\label{eq:action1}
\begin{split}
 F_{ab}&= \left[2  \log(\left\langle J \xi_{a b}, Z_{a b}\right\rangle\left\langle Z_{b a}, \xi_{b a}\right\rangle)\right.\\
 &-\left.(1+\ii\gamma) \log\left\langle Z_{a b}, Z_{a b}\right\rangle\right.\\
 &-\left.(1-\ii\gamma)  \log\left\langle Z_{b a}, Z_{b a}\right\rangle\right],\ \ a>b.
\end{split}
\ee
Here, $Z_{ab}=g_{a}^\dagger z_{ab}$; 
$\xi_{ab}$, $\zeta_{0}^{ab}$, ${J_0}_{ab}$ and  $\alpha^{(ab)(cd)}$ are the parameters given by the boundary state; $g_{a}$ and $z_{ab}$ are variables to be integrated; $J_{ab}$ are spin variables to be summed up.
In addition, we introduce a scale factor $\lambda$, such that $J_{ab}=\lambda j_{ab}$, ${J_0}_{ab}=\lambda {j_0}_{ab}$.
    
We adopt the $4$-simplex geometry used in [cite] to generate the boundary state. The five vertices of this $4$-simplex are 
\[
  \begin{split}
    P_1=(0, 0, 0, 0),\ 
    P_2=(0, 0, 0, -2 \sqrt{5}/3^{1/4}&),\ 
    P_3=(0, 0, -3^{1/4} \sqrt{5},-3^{1/4} \sqrt{5}),\\
    P_4=(0, -2 \sqrt{10}/3^{3/4}, &-\sqrt{5}/3^{3/4}, -\sqrt{5}/3^{1/4}),\\ 
    P_5=(-3^{-1/4} 10^{-1/2}, -\sqrt{5/2}&/3^{3/4}, -\sqrt{5}/3^
  {3/4}, -\sqrt{5}/3^{1/4}). 
  \end{split}
\]
Then, the $4$-normal vectors of the tetrahedra are
\begin{equation}\label{eq:4Normals}
\begin{split}
N_1=&\left(-1,0,0,0\right),\ 
N_2=\left(\frac{5}{\sqrt{22}},\sqrt{\frac{3}{22}},0,0\right),\ 
N_3=\left(\frac{5}{\sqrt{22}},-\frac{1}{\sqrt{66}},\frac{2}{\sqrt{33}},0\right),\\ 
N_4&=\left(\frac{5}{\sqrt{22}},-\frac{1}{\sqrt{66}},-\frac{1}{\sqrt{33}},\frac{1}{\sqrt{11}}\right),\ 
N_5=\left(\frac{5}{\sqrt{22}},-\frac{1}{\sqrt{66}},-\frac{1}{\sqrt{33}},-\frac{1}{\sqrt{11}}\right).
\end{split} 
\end{equation}

Table \ref{tab:facearea} lists all the ten $j_0$s. The spinors $\xi_{ba}$ and $J\xi_{ab}$ are related to the 3-normal vectors $\vec{n}_{ba}$ and $-\vec{n}_{ab}$ respectively by $\vec{n}_{ba}=\langle \xi_{ba}|\vec{\sigma}|\xi_{ba}\rangle$ and $-\vec{n}_{ab}=\langle J\xi_{ab}|\vec{\sigma}|J\xi_{ab}\rangle$. Table \ref{tab:3dNormal} (\ref{tab:xi}) records all the $3$-normal ($4$-normal) vectors of the $4$-simplex.
\begin{table}[h]
	\centering\caption{Each cell shows the area of the face shared by line number tetrahedra and column number tetrahedra.}\label{tab:facearea}
	\small
	\setlength{\tabcolsep}{0.8mm}
	\begin{tabular}{|c|c|c|c|c|}
    \hline
		\diagbox{\small{a}}{ $j_0{}_{ab}$}{\small{b}}&2&3&4&5\\
		\hline
		1&5&5&5&5\\
		\hline
		2&\diagbox{}{}&2&2&2\\
		\hline
		3&\diagbox{}{}&\diagbox{}{}&2&2\\
		\hline
		4&\diagbox{}{}&\diagbox{}{}&\diagbox{}{}&2\\
		\hline
	\end{tabular}
\end{table} 
\begin{table}[h]
	\centering\caption{Each cell shows the 3-dimensional normal vector of the face shared by line number tetrahedra and column number tetrahedra.}\label{tab:3dNormal}
	\setlength{\tabcolsep}{1.2mm}
	\vspace{1mm}
\resizebox{\textwidth}{15mm}{
  \begin{tabular}{|c|c|c|c|c|c|}
    \hline
		\diagbox{\small{a}}{normal $\vec{n}_{ab}$ }{\small{b}}&1&2&3&4&5\\
		\hline
		1&\diagbox{}{}&(1,0,0)&(-0.33,0.94,0)&(-0.33,-0.47,0.82)&(-0.33,-0.47,-0.82)\\
		\hline
		2&(-1,0,0)&\diagbox{}{}&(0.83,0.55,0)&(0.83,-0.28,0.48)&(0.83,-0.28,-0.48)\\
		\hline
		3&(0.33,-0.94,0)&(0.24,0.97,0)&\diagbox{}{}&(-0.54,0.69,0.48)&(-0.54,0.69,-0.48)\\
		\hline
		4&(0.33,0.47,-0.82)&(0.24,-0.48,0.84)&(-0.54,0.068,0.84)&\diagbox{}{}&(-0.54,-0.76,0.36)\\
		\hline
		5&(0.33,0.47,0.82)&(0.24,-0.48,-0.84)&(-0.54,0.068,-0.84)&(-0.54,-0.76,-0.36)&\diagbox{}{}\\
		\hline
	\end{tabular}
}
\end{table} 

\begin{table}[h]
  \centering\caption{Each cell indicates a spinor $\xi_{ab}$ corresponding to a 3-normal of a tetrahedron.}\label{tab:xi}
\footnotesize
\resizebox{\textwidth}{15mm}{
\begin{tabular}{|c|c|c|c|c|c|}
  \hline
  \diagbox{\small{a}}{$\ket{\xi_{ab}}$ }{\small{b}}&1&2&3&4&5\\
  \hline
  1&\diagbox{}{}&(0.71,0.71)&(0.71,-0.24+0.67i)&(0.95,-0.17-0.25i)&(0.30,-0.55-0.78i)\\
  \hline
  2&(0.71,-0.71)&\diagbox{}{}&(0.71,0.59+0.39i)&(0.86,0.48-0.16i)&(0.51,0.82-0.27i)\\
  \hline
  3&(0.71,0.24-0.67i)&(0.71,0.17+0.69i)&\diagbox{}{}&(0.86, -0.31+0.40i)&(0.51,-0.53+0.68i)\\
  \hline
  4&(0.30,0.55+0.78i)&(0.96,0.13-0.25i)&(0.96,-0.28+0.035i)&\diagbox{}{}&(0.83,-0.33-0.46i)\\
  \hline
  5&(0.95,0.17+0.25i)&(0.28,0.43-0.86i)&(0.28,-0.95+0.12i)&(0.57,-0.48-0.67i)&\diagbox{}{}\\
  \hline
\end{tabular}
}
\end{table} 

The matrix $\alpha^{(ab)(cd)}$ must have a positive definite real part, and  
\[
  \alpha^{(ab)(cd)}=\alpha_1P_0^{(ab)(cd)}+\alpha_2P_1^{(ab)(cd)}+\alpha_3P_2^{(ab)(cd)},
\]
where $\alpha_1,\alpha_2,\alpha_3$ are free parameters. The basis $P_k^{(ab)(cd)}\ (k=0\cdots2)$ are defined as
\begin{itemize}
  \item $P_0^{(ab)(cd)}=1$ if $(ab)=(cd)$ and zero otherwise;
  \item $P_1^{(ab)(cd)}=1$ if $a=c,\ b\neq d$ and zero otherwise;
  \item $P_2^{(ab)(cd)}=1$ if $(ab)\neq(cd)$ and zero otherwise.
\end{itemize}

In this paper, we set $\alpha_1=7.8816/\gamma$, $\alpha_2=0.1224/\gamma$, and $\alpha_3=1.4814/\gamma$.
The choice of $\alpha$ does not affect the application of our algorithm. 

The parameters $\zeta_0^{ab}$, whose values are given in Table \ref{tab:zeta}, are related to the dihedral angles between the 4-normal vectors \eqref{eq:4Normals}. One can find the way to determine $\zeta_0^{ab}$ in \cite{PhysRevD.103.084026}.
\begin{table}[h]
	\centering\caption{The table of $\zeta^{ab}_0$}\label{tab:zeta}
	\small
	\setlength{\tabcolsep}{0.8mm}
  \begin{tabular}{|c|c|c|c|c|c|}
    \hline
		\diagbox{\small{a}}{ $\zeta^{ab}_0$ }{\small{b}}&2&3&4&5\\
		\hline
		1&-3.14+0.36$\gamma$&0.68+0.36$\gamma$&5.05+0.36$\gamma$&5.05+0.36$\gamma$\\
		\hline
		2&\diagbox{}{}&5.05-0.59$\gamma$&-5.93-0.59$\gamma$&-3.20-0.59$\gamma$\\
		\hline
		3&\diagbox{}{}&\diagbox{}{}&-2.81-0.59$\gamma$&-5.54-0.59$\gamma$\\
		\hline
		4&\diagbox{}{}&\diagbox{}{}&\diagbox{}{}&-4.37-0.59$\gamma$\\
		\hline
	\end{tabular}
\end{table} 

By Poisson re-summation, the summation $\sum_{a>b}$ can be approximated by the integral $\int dj$  \cite{PhysRevD.103.084026} when the $\lambda$ is large.
Thus, the action and the partition function read
\be\label{eq:actiontot}
S_{tot}= \ii \lambda\sum_{ab} \zeta_0^{ab}( j_{ab}-{j_{0}}_{ab})
+\lambda\sum_{ab,cd}\alpha^{(ab)(cd)} \frac{ j_{ab}-{ j_{0}}_{ab}}{\sqrt{{ j_{0}}_{ab}}}\frac{ j_{cd}-{ j_{0}}_{cd}}{\sqrt{{ j_{0}}_{cd}}} - \sum_{a>b}\lambda j_{ab}F_{ab},
\ee
and 
\be
Z= \int \prod_{a}  \dd g_{a} \prod_{a>b} \dd j_{ab} \dd \tilde{z}_{ab}  \mathbf{d}_{\lambda j_{ab}} \e^{S_{tot}},
\ee
which has the same form as \eqref{eq:genform}. 

In our computation, we set $\gamma =0.1$ and $\lambda= 50$. 
\subsection{Pre-treatments}
To apply our saddle point finder, we apply the following pre-treatments:
\begin{itemize}
    \item
    Fix the $\SLDC$ gauge by fixing $g_1$ to be identity. 
    \item 
    Parameterize the variables $g_{a}$, $j_{ab}$, and $z_{ab}$. In the single $4$-simplex case, all the tetrahedra are boundary tetrahedra. We parameterize $g_2$ to $g_5$ as in \eqref{eq:slp}. Each $j_{ab}$ is a real variable. $z_{ab}$ are parameterized in the form \eqref{eq:spinorp}. Hence, the total action depends on $54$ real variables.  
    \item
    The works \cite{propagator,propagator1,propagator2,propagator3,PhysRevD.103.084026,Dona:2019dkf,Han:2020fil} poined out that the action \eqref{eq:actiontot} has a saddle point $s_0$ with geometric interpretation. At the saddle point $s_0$, $j_{ab}={j_0}_{ab}$, and Table \ref{tab:ga} (\ref{tab:zv}) records the values of $g_a$ (${z}_{ab}$). Using \eqref{eq:cepar}, we shift the origin of the $54$-dimensional real variables space to the saddle point $s_0$.
    \item
    The analytic continuation of the action turns all the real variables complex. We denote the analytically continued action as $\tilde{S}_{tot}$ and the analytically continued $g_{a}$, $g^\dagger_{a}$, $z_{ab}$, and conjugate $z_{ab}$ by ${\bar{g}}_{a}$, ${\bar{g}'}_{a}$, $\bar{z}_{ab}$, and ${\bar{z}'}_{ab}$. The $s_0$ is also the saddle point of $\tilde{S}_{tot}$. In $\mathbb{R}^{54}$, $|\partial_\mu \tilde{S}_{tot}|$ takes the minimal value $0$ at $s_0$. Thus, we can choose the $108$-ball centered at $s_0$ with radius $10$ as the workplace of the saddle point finder. In the subspace $\mathbb{R}^{54}$, we randomly choose $200$ points as the initial points of the coarse finder.
\end{itemize}
\begin{table}[h]
	\centering\caption{Each cell of the table is the critical point of $g_a$.}\label{tab:ga}
	\scriptsize
	\setlength{\tabcolsep}{0.5mm}
	\begin{tabular}{|c|c|c|c|c|c|}
		\hline
		\small{a}&1&2&3&4&5\\
		\hline
		${g_{0}}{}_{a}$ &$\left(\begin{matrix}
		1&0\\	0&1
		\end{matrix}\right)$&$\left(\begin{matrix}
		0.18 \ii&1.01 \ii\\	1.01 \ii&0.18 \ii
		\end{matrix}\right)$&$\left(\begin{matrix}
		0.18 \ii&0.96-0.34 \ii\\	-0.96-0.34 \ii&0.18 \ii
		\end{matrix}\right)$&$\left(\begin{matrix}
		1.01 \ii&-0.48-0.34 \ii\\	0.48-0.34 \ii&-0.65 \ii
		\end{matrix}\right)$&$\left(\begin{matrix}
		-0.65 \ii&-0.48-0.34 \ii\\	0.48-0.34 \ii&1.01 \ii
		\end{matrix}\right)$\\
		\hline
	\end{tabular}
\end{table} 
\begin{table}[htbp]
  \centering\caption{Each cell indicates a spinor $z_{ab}$.}\label{tab:zv}
\footnotesize
\setlength{\tabcolsep}{0.8mm}
\begin{tabular}{|c|c|c|c|c|c|}
  \hline
  \diagbox{\small{a}}{$\ket{z_0{}_{ab}}$ }{\small{b}}&1&2&3&4&5\\
  \hline
  1&\diagbox{}{}&(1,1)&(1,-0.333+0.942i)&(1,-0.184-0.259i)&(1,-1.817-2.569i)\\
  \hline
  2&(1,1)&\diagbox{}{}&(1,0.685-0.729i)&(1,1.857+0.989i)&(1,0.420+0.223i)\\
  \hline
  3&(1,0.333-0.943i)&(1,0.685-0.729i)&\diagbox{}{}&(1, 0.313+2.080i)&(1,0.071+0.470i)\\
  \hline
  4&(1,-0.184-0.259i)&(1, 1.857+0.989i)&(1,0.313+2.080i)&\diagbox{}{}&(1, 0.058+0.082i)\\
  \hline
  5&(1,-1.817-2.569i)&(1,0.420+0.223i)&(1,0.071+0.470i)&(1,  0.058+0.082i)&\diagbox{}{}\\
  \hline
\end{tabular}
\end{table}

\subsection{Results}
Other than $s_0$, our saddle-point finder finds two more complex saddle points $s_1$ and $s_2$.
At $s_1$, Tables \ref{tab:facearea1} to \ref{tab:czv1} show all the $j_{ab}$, $\bar{g}_a$, ${\bar{g}'}_a$, $\bar{z}_{ab}$, and ${\bar{z}'}_{ab}$ respectively.
At $s_2$, Tables \ref{tab:facearea2} to \ref{tab:czv2} show all the $j_{ab}$, $\bar{g}_a$, ${\bar{g}'}_a$, $\bar{z}_{ab}$, and ${\bar{z}'}_{ab}$ respectively.
The values of the action $\tilde{S}_{tot}$ at $s_0$, $s_1$, and $s_2$ are $0+138.037 \ii$, $-0.334705 + 138.179 \ii$, and $-0.551927 + 137.624 \ii$.
The real parts indicate that by contribution to the partition function, $s_0>s_1>s_2$.
\subsection{Geometrical interpretations}
The work \cite{PhysRevD.105.024012} shows that the bivectors generated by group variables $g_a$ and spinors $z_{ab}$ and $\xi_{ab}$ encode the geometric interpretation of a complex saddle point. 
Let
\be
\begin{aligned}
\chi_{ab}^{\prime} &=\frac{\mathrm{i} \gamma+\kappa_{ab} }{\mathrm{i} \gamma-1} \frac{\bar{Z}_{ab}^{\prime}}{\bar{Z}_{ab}^{\prime} \bar{Z}_{ab}}-\frac{\kappa_{ab}+1}{\mathrm{i} \gamma-1} \frac{\xi_{ab}^{\dagger}}{\xi_{ab}^{\dagger}  \bar{Z}_{ab}}, \\
\chi_{ab} &=\frac{\mathrm{i} \gamma+\kappa_{ab} }{\mathrm{i} \gamma+1} \frac{\bar{Z}_{ab}}{\bar{Z}_{ab}^{\prime}  \bar{Z}_{ab}}-\frac{ \kappa_{ab}-1}{\mathrm{i} \gamma+1} \frac{\xi_{ab}}{\bar{Z}_{ab}^{\prime}\xi_{ab}},
\end{aligned}
\ee
where 
\[
\bar{Z}_{ab}^{\prime}=\bar{z}^{\prime}_{ab}\bar{g}_b,\,\ 
\bar{Z}_{ab}=\bar{g}^\prime_a\bar{z}_{ab},
\]
and
\[
k_{ab}=\left\{
\begin{matrix}
1,&&\,a>b,\\
-1,&&\,a<b\ \ \ .\\
\end{matrix}
\right.
\]
Two traceless simple bivectors of the face $ab$ are defined by
\be
B^{+}_{ab}= \chi_{ab} \otimes \bar{Z}^\prime_{ab}-\half \mathbf{1},
\ee
\be
B^{-}_{ab}=\bar{Z}_{ab} \otimes \chi_{ab}^{\prime}-\half \mathbf{1}.
\ee
The $4$-dimensional bivectors ${B_{\pm}}_{ab}^{IJ}$ of the face $ab$ are the spin-1 representations of $B^{\pm}_{ab}$.
Namely, 
\[
{B_{\pm}}_{ab}^{IJ}=
\left(\begin{array}{cccc}
0 & K_{\pm}^{1} & K_{\pm}^{2} & K_{\pm}^{3} \\
-K_{\pm}^{1} & 0 & J_{\pm}^{3} & -J_{\pm}^{2} \\
-K_{\pm}^{2} & -J_{\pm}^{3} & 0 & J_{\pm}^{1} \\
-K_{\pm}^{3} & J_{\pm}^{2} & -J_{\pm}^{1} & 0
\end{array}\right),
\]
where 
\[
K^{i}_{\pm}+\mathrm{i} J^{i}_{\pm}=\operatorname{Tr}\left({B^{\pm}}_{ab} \sigma^{i}\right),
\]
and $\sigma_i$ are Pauli matrices.
For each tetrahedron $a$, the closure condition reads
\be\label{eq:cl}
\begin{split}
 \sum_{b \in \{1\cdots 5\}\setminus a} j_{ab} \kappa_{ab} B_{ab}^{-}=0,\\
 \sum_{b \in \{1\cdots 5\}\setminus a} j_{ab} \kappa_{ab} B_{ab}^{+}=0.
\end{split}
\ee
For each face $ab$, the parallel condition reads
\be\label{eq:pl}
\left(\bar{g}_{a}^{\prime}\right)^{-1} B_{ab}^{-} \bar{g}_{a}^{\prime}=-\left(\bar{g}_{b}^{\prime}\right)^{-1} B_{ba}^{-} \bar{g}_{b}^{\prime}, \quad \bar{g}_{a} B_{ab}^{+}\left(\bar{g}_{a}\right)^{-1}=-\bar{g}_{b} B_{ba}^{+}\left(\bar{g}_{b}\right)^{-1}.
\ee
Saddle point $s_1$ meets \eqref{eq:cl} and \eqref{eq:pl}, while
$s_2$ meets \eqref{eq:cl} and \eqref{eq:pl} up to an error of $10^{-5}$.
At either $s_1$ or $s_2$, however, for each tetrahedron $a$, one cannot find its $4$-dimensional normal $N_I$ that meets the condition
\[
\forall b \in \{1\cdots 5\}\setminus a,\,\ {B_{\pm}}_{ab}^{IJ}N_J=\mathbf{0}.
\]
Thus, both $s_1$ and $s_2$ are saddle points with Lorentzian $SO(1,3)$ bivector geometry\footnote{The value of $B_{\pm}$ bivectors can be found in our program}. 
\section{Application: Saddle points in the $\Delta_3$ EPRL spin foam model}
\subsection{The action}
\begin{figure}[htbp]
\centering\includegraphics[width=0.75\linewidth]{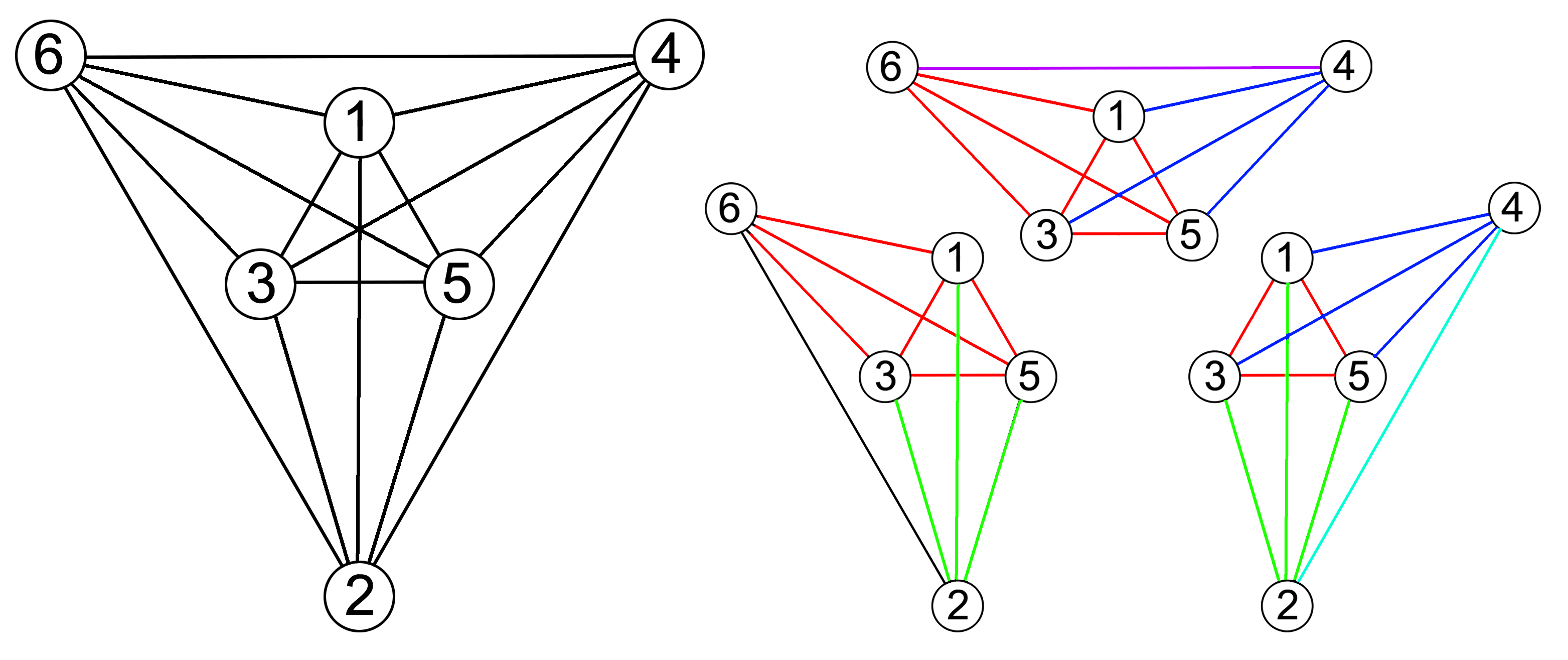}
\caption{The left side figure indicates the topological structure of the $\Delta_3$. A $\Delta_3$ consists $6$ vertices and the edges connecting every two vertices. The right side figure shows that the $\Delta_3$ can be decomposed into three $4$-simplices.}\label{fig:d3ps}
\end{figure}
The simplicial complex $\mathcal{K}$ considered in this section consists of three $4$-simplices as in Fig. \ref{fig:d3ps}. 
We follow the convention in \cite{dona2020numerical,han2021complex} to call this $\mathcal{K}$ as $\Delta_3$.
As shown in Fig. \ref{fig:d3ps}, we number the vertices of the $\Delta_3$ from $1$ to $6$. 
Each $4$-simplex is labeled by a single index $a$. We let $a=6$ for the $4$-simplex $12345$, $a=4$ for the $4$-simplex $12356$, and $a=2$ for the $4$-simplex $13456$. 
In $4$-simplex $a$, the number pair $ab$ labels the tetrahedron whose vertices belong to the set $\{1,\cdots,6\}\setminus\{a,b\}$.
For example, five tetrahedra belonging to $4$-simplex $6$ are labeled by $61,\,62,\,63,\,64,\,65$. 
The face shared by $ab$ and $ac$ is labeled by the triple $abc$. 
The faces in the $\Delta_3$ are classified into three types.
\begin{enumerate}
    \item Type I consists the faces belonging only to a single $4$-simplex. The labels of this type of the faces form the set
    \be
    F1=\{ abc|\ a\in\{2,4,6\},\  b,c\in\{1,3,5\},\ \text{and}\ b\neq c\}.
    \ee
    \item Type II faces belong to the tetrahedra shared by two $4$-simplices. The labels of the type II face form the set
    \be
    F2=\{ abc|\ (a,b\in\{2,4,6\},\ c\in\{2,4,6\})\ \text{or}\ (a,c\in\{2,4,6\},\ b\in\{2,4,6\})\}.
    \ee
    \item Type III faces are shared by three $4$-simplices. Type III face's labels form the set
    \be
     F3=\{ abc|\ a,b,c\in\{2,4,6\} \}.
    \ee
\end{enumerate}

The partition function of this $\Delta_3$ spin foam is 
\be\label{eq:amp2}
Z=\sum_{\{J_{abc} | abc\in F3\}}\prod_{{abc}} \mathbf{d}_{J_{abc}} \int [\dd X] \e^{S(X,J_{abc})},
\ee
where 
\be\label{eq:setup2}
\left\{
\begin{split}
    \mathbf{d}_{J_{abc}}&=2J_{abc}+1;\\
    X&\equiv (g_{ab},z_{abc},J_{abc}^I);\\
    \dd X & \equiv (\dd g_{ab}, \dd \tilde{z}_{abc});\\
    T&\equiv (\xi_{abc},J_{abc}^B),\, abc\notin F3.\\
\end{split}
\right.
\ee
In \eqref{eq:setup2}, we denote $\{J_{abc}| abc\in F3 \}$ as $J^I_{abc}$ and $\{J_{abc}| abc\notin F3 \}$ as $J^{B}_{abc}$.
In contrast to \eqref{eq:setup1}, all the internal $\xi^I$ have already been integrated out and thus are not the components of $X$. 

In our convention, the face $abc$ is glued to the face labeled by any permutation of $abc$, e.g., face $624$ is glued to face $642$. The spinor variables $z$  and spin variables $J$ assigned to the glued faces fulfill the following rules.
\begin{enumerate}
    \item The glued faces $abc$ and $acb$ belong to the same $4$-simplex, denoted by $a$, and share a spinor variable $z_{abc}$. 
    \item The glued faces $abc$ and $bac$ belong to two different $4$-simplices, and each has its own spinor variable, i.e., $z_{abc}\neq z_{bac}$. 
    \item Any two glued faces share a spin variable $J$. Thus, for any permutation of $abc$, denoted as $[abc]$, $J_{abc}$=$J_{[abc]}$.
\end{enumerate}
The $X$ in \eqref{eq:setup2} contains $15$ group $g$ variables, $30$ spinor $z$ variables, and $1$ spin $J$ variable.
Each boundary face is assigned with a spinor $\xi$. Therefore,
the $T$ in \eqref{eq:setup2} contains $36$ spinor $\xi$ parameters and $18$ spin $J$ parameters.

Note that all the spin variables are half-integer valued, and we have $\sum_{\{J_{abc} | abc\in F3\}}$ instead of $\int \dd J_{abc}$ in \eqref{eq:amp2}. 
In order to apply our saddle-point finder, we use Poisson summation to approximate the summation over the internal spin by the integral over continuous $J_{abc}$ in the large spin region \cite{han2021complex}. 
For convenience, we introduce a scale factor $\lambda$ of the spin variables, such that $J_{abc}=\lambda j_{abc}$. 
We apply our saddle-point finder in the case with $\lambda=50$. 
In this case, the partition function is approximated by 
\be
Z=\int [\dd X]  \dd \lambda j_{246} \prod_{{abc}} \mathbf{d}_{\lambda j_{abc}}  \e^{\lambda S(X)}.
\ee
The action $S$ contains three parts,
\be
S=S_1+S_2+S_3.
\ee
Type I, Type II, and Type III faces contribute to $S_1$, $S_2$, and $S_3$ respectively.
Let $Z_{abc}=g_{ab}^\dagger z_{abc}$, we have
\be\begin{split}
    S_1&=\sum_{abc\in f_1}\left(j_{abc} \ln \frac{\left\langle\xi_{abc}, Z_{abc}\right\rangle^{2}\left\langle Z_{acb}, \xi_{acb}\right\rangle^{2}}{\left\langle Z_{abc}, Z_{abc}\right\rangle\left\langle Z_{acb}, Z_{acb}\right\rangle} +\ii \gamma j_{abc} \ln \frac{\left\langle Z_{acb}, Z_{acb}\right\rangle}{\left\langle Z_{abc}, Z_{abc}\right\rangle}\right),
\end{split}
\ee
\be
\begin{split}
S_2&=\sum_{abc \in f_2}\left(j_{abc} \ln \frac{\left\langle\xi_{abc}, Z_{abc}\right\rangle^{2}\left\langle Z_{cba}, \xi_{cba}\right\rangle^{2}}{\left\langle Z_{abc}, Z_{abc}\right\rangle\left\langle Z_{cba}, Z_{cba}\right\rangle}\frac{\left\langle Z_{acb}, Z_{cab}\right\rangle^{2}}{\left\langle Z_{cab}, Z_{cab}\right\rangle\left\langle Z_{acb}, Z_{acb}\right\rangle}\right.\\
&\left.+\ii \gamma j_{abc} \ln \frac{\left\langle Z_{cba}, Z_{cba}\right\rangle}{\left\langle Z_{abc}, Z_{abc}\right\rangle}\frac{\left\langle Z_{acb}, Z_{acb}\right\rangle}{\left\langle Z_{cab}, Z_{cab}\right\rangle}\right),
\end{split}
\ee
and 
\be
\begin{split}
S_3 &= j_{246}\left[\ln \frac{\left\langle Z_{6 4 2}, Z_{4 6 2}\right\rangle^{2}}{\left\langle Z_{4 6 2}, Z_{4 6 2}\right\rangle\left\langle Z_{6 4 2}, Z_{6 4 2}\right\rangle}+\ii \gamma \ln \frac{\left\langle Z_{6 4 2}, Z_{6 4 2}\right\rangle}{\left\langle Z_{4 6 2}, Z_{4 6 2}\right\rangle}\right]\\
&+j_{246}\left[\ln \frac{\left\langle Z_{4 2 6}, Z_{2 4 6}\right\rangle^{2}}{\left\langle Z_{2 4 6}, Z_{2 4 6}\right\rangle\left\langle Z_{4 2 6}, Z_{4 2 6}\right\rangle}+\ii \gamma \ln \frac{\left\langle Z_{4 2 6}, Z_{4 2 6}\right\rangle}{\left\langle Z_{2 4 6}, Z_{2 4 6}\right\rangle}\right]\\
&+j_{246}\left[\ln \frac{\left\langle Z_{2 6 4}, Z_{6 2 4}\right\rangle^{2}}{\left\langle Z_{6 2 4}, Z_{6 2 4}\right\rangle\left\langle Z_{2 6 4}, Z_{2 6 4}\right\rangle}+\ii \gamma \ln \frac{\left\langle Z_{2 6 4}, Z_{2 6 4}\right\rangle}{\left\langle Z_{6 2 4}, Z_{6 2 4}\right\rangle}\right],\\
\end{split}
\ee
where
\[
f_1=\{
635,413,453,451,235,251,231
\},
\]
and
\[
f_2=
\{
216,416,436,632,652,654,432,214,254,615
\}.
\]
The $\gamma$ above is the Immirzi parameter. In our work, we set $\gamma=0.2$.
We remark that the order of the numbers of each element in $f_1$ and $f_2$ and the explicit form of $S_3$ depend on the orientation of the $\Delta_3$ complex. 

The parameters $\xi_{vef}$ and $j^B_{abc}$ are given by the simplicial geometry of the $\Delta_3$. This geometry is determined by the $15$ edge-lengths. 
Here, we denote each edge by $ab$, with $a$ and $b$ the ends of the edge.
\begin{table}[htbp]
  \centering\caption{Edge-lengths in $4$-simplex $6$}\label{tab:s6}
\footnotesize
\begin{tabular}{|c|c|c|c|c|c|}
  \hline
  \diagbox{\small{a}}{$l_{ab}$ }{\small{b}}&1&2&3&4&5\\
  \hline
  1&\diagbox{}{}&$\sqrt{11.547}$&$\sqrt{11.547}$&$\sqrt{4.272}$&$\sqrt{11.547}$\\
  \hline
  2&\diagbox{}{}&\diagbox{}{}&$\sqrt{11.547}$&$\sqrt{4.272}$&$\sqrt{11.547}$\\
  \hline
  3&\diagbox{}{}&\diagbox{}{}&\diagbox{}{}&$\sqrt{4.272}$&$\sqrt{11.547}$\\
  \hline
  4&\diagbox{}{}&\diagbox{}{}&\diagbox{}{}&\diagbox{}{}&$\sqrt{4.272}$\\
  \hline
\end{tabular}
\end{table}
Since edges $15$, $35$, and $13$ are shared by all three $4$-simplices, edges $21$, $23$, and $25$ are shared by $4$-simplices $4$ and $6$, and edges $41$, $43$, and $45$ are shared by $4$-simplices $2$ and $6$, one only needs to set the length of the edges $61$, $62$, $63$, $62$, and $64$ to fix the $\Delta_3$. 
In the case with cylindrical symmetry \cite{hanPI},
\[
l_{61}=l_{63}=l_{65}=l_1,\,
l_{62}=l_2,\,
l_{64}=l_3.
\]
We set $l_1=\sqrt{12.8421}$, $l_2=\sqrt{33.3319}$, and  $l_3=\sqrt{17.1054}$. 
The $4$-normal vectors of tetrahedra $64$ and $62$ are 
\[
N_{64}=(-1,0,0,0),\, N_{62}=(1.066,0.369,0,0);
\]
the $4$-normal vectors of tetrahedra $46$ and $42$ are 
\[
N_{46}=(1,0,0,0),\, N_{42}=(1,-0.00173,0,0);
\]
the $4$-normal vectors of tetrahedra $26$ and $24$ are
\[
N_{26}=(-1.066,-0.369,0,0), N_{24}=(-1.473,1.082,0,0).
\]

In each $4$-simplex, the inner product of the $4$-normal vectors of two tetrahedra defines the dihedral angle on the common face of the two tetrahedra. For example, in $4$-simplex $6$, the dihedral angle $\theta^{6}_{642}$ on face $642$ satisfies
\[
\cosh{\left(\theta^{6}_{642}\right)}= \eta_{ij}N_{64}^i N_{62}^j.
\]

With our given edge-lengths, $\theta^{6}_{642}=0.361$, $\theta^{4}_{462}=0.00172$, and $\theta^{2}_{246}=1.2995$. The deficit angle $\theta^D_{246}$ hinged on face $246$ depends on the orientation of the $\Delta_3$ and can take one of the following $8$ values
\begin{eqnarray}
\theta^D_{246}&=&0.3614-0.001726-1.300=-0.9399,\label{eq:d1}\\
\theta^D_{246}&=&0.3614-0.001726+1.300=1.659,\label{eq:d2}\\
\theta^D_{246}&=&0.3614+0.001726-1.300=-0.9364,\label{eq:d3}\\
\theta^D_{246}&=&-0.3614-0.001726-1.300=-1.662,\label{eq:d4}\\
\theta^D_{246}&=&0.3614+0.001726+1.300=1.662,\label{eq:d5}\\
\theta^D_{246}&=&-0.3614-0.001726+1.300=0.9364,\label{eq:d6}\\
\theta^D_{246}&=&-0.3614+0.001726-1.300=-1.659,\label{eq:d7}\\
\theta^D_{246}&=&-0.3614+0.001726+1.300=0.9399.\label{eq:d8}
\end{eqnarray}

With these edge-lengths, one can compute the $3$-normal vector of each face in $\Delta_3$, and use these $3$-normal vectors to build the $\xi$ and the $j^B$. Tables \ref{tab:xv1} to \ref{tab:xv3} record the values of $\xi$.
\begin{table}[htbp]
  \centering\caption{Values of $\xi_{6ab}$.}\label{tab:xv1}
\footnotesize
\resizebox{\textwidth}{12mm}{
\begin{tabular}{|c|c|c|c|c|c|}
  \hline
  \diagbox{\small{a}}{$\xi_{6ab}$ }{\small{b}}&1&2&3&4&5\\
  \hline
  1&\diagbox{}{}&(0.2887,-0.9534+0.0878i)&(0.9574,-0.1667-0.2357i)&(1,0)&(0.9574,-0.1208+0.2622i)\\
  \hline
  2&(0.9574,-0.25-0.1443i)&\diagbox{}{}&(0.2887,-0.8292+0.4787i)&(0,1)&(0.2887,-0.9574i)\\
  \hline
  3&(0.2887,-0.5528-0.7817i)&(0.9574,0.1208-0.2622i)&\diagbox{}{}&(1,0)&(0.9574,-0.2875+0.02649i)\\
  \hline
  4&(0.9530,0.1749+0.2473i)&(0.7071,0.7071)&(0.3029,0.5502+0.7781i)&\diagbox{}{}&(0.7071,-0.2357+0.6667i)\\
  \hline
  5&(0.2887,0.1750+0.9413i)&(0.9574,0.2722+0.09623i)&(0.2887,0.7277-0.6222i)&(0,0.9701+0.2425i)&\diagbox{}{}\\
  \hline
\end{tabular}
}
\end{table}

\begin{table}[htbp]
  \centering\caption{Values of $\xi_{4ab}$.}\label{tab:xv2}
\footnotesize
\resizebox{\textwidth}{12mm}{
\begin{tabular}{|c|c|c|c|c|c|}
  \hline
  \diagbox{\small{a}}{$\xi_{4ab}$}{\small{b}}&1&2&3&5&6\\
  \hline
  1&\diagbox{}{}&(0.9467,0.1348-0.2926i)&(0.9530,-0.1749-0.2473i)&(0.3029,-0.9490+0.08744i)&(0,-0.6247-0.7809i)\\
  \hline
  2&(0.8096,0.5083-0.2935i)&\diagbox{}{}&(0.5870,0.7011+0.4048i)&(0.8096,0.5870i)&(1,0)\\
  \hline
  3&(0.3029,-0.5502-0.7781i)&(0.3221,-0.9427+0.08686i)&\diagbox{}{}&(0.3029,-0.3988+0.8656i)&(0,0.5145+0.8575i)\\
  \hline
  5&(0.9530,0.2302-0.1968i)&(0.9467,0.3037+0.1074i)&(0.9530,0.05535+0.2978i)&\diagbox{}{}&(1,0)\\
  \hline
  6&(0.9530,0.1749+0.2473i)&(0.7071,0.7071)&(0.3029,0.5502+0.7781i)&(0.7071,-0.2357+0.6667i)&\diagbox{}{}\\
  \hline
\end{tabular}
}
\end{table}

\begin{table}[htbp]
  \centering\caption{Values of $\xi_{2ab}$.}\label{tab:xv3}
\footnotesize
\resizebox{\textwidth}{12mm}{
\begin{tabular}{|c|c|c|c|c|c|}
  \hline
  \diagbox{\small{a}}{$\xi_{2ab}$}{\small{b}}&1&3&4&5&6\\
  \hline
  1&\diagbox{}{}&(0.9685,-0.2171-0.1220i)&(0.9590,-0.09545-0.2667i)&(0.9685,0.09038-0.2320i)&(0.9985,0.01820+0.05084i)\\
  \hline
  3&(0.2490,-0.8443-0.4745i)&\diagbox{}{}&(0.2833,-0.9393+0.1935i)&(0.2490,-0.5880+0.7696i)&(0.05400,0.9780-0.2015i)\\
  \hline
  4&(0.8096,0.5083-0.2935i)&(0.5870,0.7011+0.4048i)&\diagbox{}{}&(0.8096,0.5870i)&(1,0)\\
  \hline
  5&(0.2490,-0.9684-0.01161i)&(0.9685,0.03190+0.2469i)&(0.9590,0.2116+0.1883i)&\diagbox{}{}&(0.05400,0.7460+0.6638i)\\
  \hline
  6&(0.9574,-0.25-0.1443i)&(0.2887,-0.8292+0.4787i)&(0,1)&(0.2887,-0.9574i)&\diagbox{}{}\\
  \hline
\end{tabular}
}
\end{table}

\begin{minipage}{\textwidth}
\begin{minipage}{0.2\linewidth}
\centering
\makeatletter\def\@captype{table}\makeatother\caption{The $j_{6ab}^b$}\label{tab:arv1}
	\small
	\setlength{\tabcolsep}{0.8mm}
	\footnotesize
	\begin{tabular}{|c|c|c|c|c|c|}
    \hline
		\diagbox{\small{a}}{ $j^b_{6ab}$}{\small{b}}&1&2&3&4&5\\
		\hline
		1&\diagbox{}{}&2&2&5&2\\
		\hline
		2&2&\diagbox{}{}&2&\diagbox{}{}&2\\
		\hline
		3&2&2&\diagbox{}{}&5&2\\
		\hline
		4&5&\diagbox{}{}&5&\diagbox{}{}&5\\
		\hline
		5&2&2&2&5&\diagbox{}{}\\
		\hline
	\end{tabular}
\end{minipage}
\begin{minipage}{0.38\linewidth}
\centering
\makeatletter\def\@captype{table}\makeatother\caption{The $j_{4ab}^b$}\label{tab:arv2}
	\small
	\setlength{\tabcolsep}{0.8mm}
	\footnotesize
	\begin{tabular}{|c|c|c|c|c|c|}
    \hline
		\diagbox{\small{a}}{ $j_{4ab}^b$}{\small{b}}&1&2&3&5&6\\
		\hline
		1&\diagbox{}{}&5.361&5.663&5.663&5\\
		\hline
		2&5.361&\diagbox{}{}&5.361&5.361&\diagbox{}{}\\
		\hline
		3&5.663&5.361&\diagbox{}{}&5.663&5\\
		\hline
		5&5.663&5.361&5.663&\diagbox{}{}&5\\
		\hline
		6&5&\diagbox{}{}&5&5&\diagbox{}{}\\
		\hline
	\end{tabular}
\end{minipage}
\begin{minipage}{0.3\linewidth}
\centering
\makeatletter\def\@captype{table}\makeatother\caption{The $j_{2ab}^b$}\label{tab:arv3}
	\small
	\setlength{\tabcolsep}{0.8mm}
	\footnotesize
	\begin{tabular}{|c|c|c|c|c|c|}
    \hline
		\diagbox{\small{a}}{$j_{2ab}^b$}{\small{b}}&1&3&4&5&6\\
		\hline
		1&\diagbox{}{}&3.704&5.361&3.704&2\\
		\hline
		3&3.704&\diagbox{}{}&5.361&3.704&2\\
		\hline
		4&5.361&5.361&\diagbox{}{}&5.361&\diagbox{}{}\\
		\hline
		5&3.704&3.704&5.361&\diagbox{}{}&2\\
		\hline
		6&2&2&\diagbox{}{}&2&\diagbox{}{}\\
		\hline
	\end{tabular}
\end{minipage}
\end{minipage}

Tables \ref{tab:arv1} to \ref{tab:arv3} record the variables $j^B$.
Many values in Tables \ref{tab:arv2} and \ref{tab:arv3} are not half-integers; however, at large $\lambda$, the difference between $\lambda j^B$ and its closest half integer is negligible. Therefore, a $\lambda j^B$ can be approximately regarded as a half-integer spin variable.

\subsection{Pre-treatments}
We have the following pre-treatments.
\begin{enumerate}
    \item We fix the gauges \eqref{eq:gsldc} and \eqref{eq:gsu} by parameterizing the group variables as follows. The $\SLDC$ gauge on each $4$-simplex is fixed by restricting $g_{61}$, $g_{45}$, and $g_{23}$ to be the identity matrix. The $\Su$ gauge on each internal tetrahedron is fixed by parameterizing $g_{64}$, $g_{42}$, and $g_{26}$ as in \eqref{eq:sup}. The group variables $g_{65},\,g_{63},\,g_{43},\,g_{41},\,g_{25},\,g_{21},\,g_{62},\,g_{46},\,g_{24}$ are parameterized as in \eqref{eq:slp}
    \item All the $z$ variables are parameterized as \eqref{eq:spinorp}. All the $j$s are already real variables and hence needs no additional parameterization.   
    \item The works \cite{Barrett:2009mw,BARRETT_2010,CFsemiclassical,LowE1,HZ,HZ1,Kaminski:2017eew} pointed out that the simplicial geometry defines the critical points of the spin foam action $X_{a}=(j_a,\,z_a,\,g_a)$, such that 
    \[
    \begin{split}
        \mathrm{Re}(S(X_a))=0,\\
        \partial_{g}S|_{X_a}=0,\\
        \partial_z S|_{X_a}=0.
    \end{split}
    \]
    In our case with curvature, these critical points are not saddle points because 
    \[
    \mathrm{Im}\left(\partial_{j_{246}} S |_{X_a}\right)=\gamma\lambda\theta^D_{246}.
    \] 
   Such critical points and the points close to them can still be the initial points of our saddle-point finder. 
   
   Corresponding to the simplicial geometry with deficit angle \eqref{eq:d1}, the $g^0_{ab}$ and $z^0_{abc}$ of a critical point $X_0$ are given in Tables \ref{tab:g0a1} to \ref{tab:zv22}, and  $j^0_{246}$ is $5$. We shift the origin of the space of our real variables to $X_0$ by plugging $g^0_{ab}$ and $z^0_{abc}$ into \eqref{eq:cepar}. In this parameterization, the action $S$ depends on $124$ real variables.
   
   Seven more critical points can be found by acting parity flip operation on $X_0$.
   On each $4$-simplex, the parity flip is a transformation between two critical points $(g^0_{ab},\,z^0_{abc},\,j^0_{abc})$ and $(\tilde{g}^0_{ab},\,\tilde{z}^0_{abc},\,\tilde{j}^0_{abc})$, where
   \[
   \tilde{g}^0_{ab}=(g^{0\dagger}_{ab})^{-1},
   \]
   \[
   \tilde{z}^0_{abc}=\frac{g^0_{ab}g^{0\dagger}_{ab}z^0_{abc}}{||g^{0\dagger}_{ab}z^0_{abc}||^2},
   \]
   and 
   \[
   \tilde{j}^0_{abc}=j^0_{abc}.
   \]
   In $\Delta_3$, including the identity, there are $2^3$ different ways of parity flipping. Acting these flippings on $X_0$ results in $7$ more critical points.
   These critical points corresponding to the simplicial geometries with deficit angles \eqref{eq:d2} to \eqref{eq:d8}. Using the technique introduced in Appendix C, one can find the coordinates of those critical points in our parameterization.  
    
    \item Similar to the single 4-simplex case, the analytic continuation of the action changes all the real variables into complex. We denote the analytically continued action as $\tilde{S}$ and the analytically continued $g_{ab}$, $g_{ab}^\dagger$, $z_{abc}$, and conjugate $z_{abc}$ as ${\bar{g}}_{ab}$, ${\bar{g}'}_{ab}$, $\bar{z}_{abc}$ and ${\bar{z}'}_{abc}$. The $j_{246}$ is analytically continued as $\tilde{j}_{246}$. 
    
    The distances between the $7$ additional critical points and $X_0$ are smaller than $21$. Therefore, We choose the $248$-ball centered at $X_0$ with radius $21$ as the workplace of the saddle point finder. In the subspace $\mathbb{R}^{124}$, we randomly choose $1600$ points closing to the critical points, and feed them to our saddle-point finder.

\end{enumerate}
\subsection{Results}
    The saddle-point finder finds $112$ points. The maximal\ value of $\partial_{\mu}\tilde{S}$ at these points is $8.91\times10^{-12}$, so we can safely consider all $112$ points as the saddle points of the $\tilde{S}$. We store the exact values of the $(\tilde{S}, \bar{g}_{ab},\, {\bar{g}'}_{ab},\, \bar{z}_{abc},\, {\bar{z}'}_{abc},\, \tilde{j}_{246})$ at each saddle points in \cite{hzcgit2}. We can compute the real part of $\tilde S$. In our computation, we find $44$ saddle points have positive real part of the action. By \cite{PhysRevD.103.084026}, we know that the saddle points attached to the Lefschetz thimbles have negative real part of the action. Hence, those $44$ saddle points with positive real part of the action are attached with the anti-thimbles and do not contribute to the partition function. The other $68$ saddle points with negative real part of the action contribute to the partition function, and each point's contribution can be estimated by its real part of the action.
    
\subsection{Geometrical Interpretations}
    As we mentioned, the geometrical interpretation of the saddle points is encoded in the bivectors. For each face $abc$, two bivectors can be defined 
    \be
    B^{+}_{abc}= \chi_{abc} \otimes \bar{Z}^\prime_{abc}-\half \mathbf{1},
    \ee
    \be
    B^{-}_{abc}=\bar{Z}_{abc} \otimes \chi_{abc}^{\prime}-\half \mathbf{1}.
    \ee
    When $b\in\{1,\,3,\,5\}$, the face $abc$ is a boundary face, and its $\chi_{abc}^{\prime}$ and $\chi_{abc}$ read
    \be
    \begin{aligned}
    \chi_{abc}^{\prime} &=\frac{\mathrm{i} \gamma+\kappa_{abc} }{\mathrm{i} \gamma-1} \frac{\bar{Z}_{abc}^{\prime}}{\bar{Z}_{abc}^{\prime} \bar{Z}_{abc}}-\frac{\kappa_{abc}+1}{\mathrm{i} \gamma-1} \frac{\xi_{abc}^{\dagger}}{\xi_{abc}^{\dagger}  \bar{Z}_{abc}}, \\
    \chi_{abc} &=\frac{\mathrm{i} \gamma+\kappa_{abc} }{\mathrm{i} \gamma+1} \frac{\bar{Z}_{abc}}{\bar{Z}_{abc}^{\prime}  \bar{Z}_{abc}}-\frac{ \kappa_{abc}-1}{\mathrm{i} \gamma+1} \frac{\xi_{abc}}{\bar{Z}_{abc}^{\prime}  \xi_{abc}}.
    \end{aligned}
    \ee
    When $b\in\{2,\,4,\,6\}$, the face $abc$ is a bulk face, and
    \be
    \begin{aligned}
    \chi_{abc}^{\prime} &=\frac{\mathrm{i} \gamma+\kappa_{abc} }{\mathrm{i} \gamma-1} \frac{\bar{Z}_{abc}^{\prime}}{\bar{Z}_{abc}^{\prime} \bar{Z}_{abc}}-\frac{\kappa_{abc}+1}{\mathrm{i} \gamma-1} \frac{{\bar{Z}'}_{abc}}{{\bar{Z}'}_{abc}  \bar{Z}_{abc}}, \\
    \chi_{abc} &=\frac{\mathrm{i} \gamma+\kappa_{abc} }{\mathrm{i} \gamma+1} \frac{\bar{Z}_{abc}}{\bar{Z}_{abc}^{\prime}  \bar{Z}_{abc}}-\frac{ \kappa_{abc}-1}{\mathrm{i} \gamma+1} \frac{\bar{Z}_{abc}}{\bar{Z}_{abc}^{\prime}  \bar{Z}_{abc}}.
    \end{aligned}
    \ee
    The $\kappa_{abc}$ depends the orientation of the $\Delta_3$. Namely, $\kappa_{abc} = -1$ for the faces $612,\,614,\,623,\,\allowbreak 625,\,631,\, 634,\,651,\,653,\,412,\,415,\,423,\,426,\,431,\,435,\,451,\,456,\,461,\,463,\,213,\,215,\,234,\,236,\,\allowbreak 241,\,\allowbreak 245,\,\allowbreak 253,\,\allowbreak 256,\,\allowbreak 361,\,\text{and}\,\, 264$, otherwise $\kappa_{abc}=1$.
    For each tetrahedron $ab$, the closure condition is given by 
    \be\label{eq:cl1}
    \begin{split}
    \sum_{c \in \{1\cdots 6\}\setminus\{ab\}} j_{abc} \kappa_{abc} B_{abc}^{-}=0,\\
    \sum_{c \in \{1\cdots 6\}\setminus\{ab\}} j_{abc} \kappa_{abc} B_{abc}^{+}=0.
    \end{split}
    \ee
    For each face $abc$, the parallel condition reads
    \be\label{eq:pl1}
    \left(\bar{g}_{ab}^{\prime}\right)^{-1} B_{abc}^{-} \bar{g}_{ab}^{\prime}=-\left(\bar{g}_{ac}^{\prime}\right)^{-1} B_{acb}^{-} \bar{g}_{ac}^{\prime}, \quad \bar{g}_{ab} B_{abc}^{+}\left(\bar{g}_{ab}\right)^{-1}=-\bar{g}_{ac} B_{acb}^{+}\left(\bar{g}_{ac}\right)^{-1}.
    \ee 
    
    All the saddle points satisfy the closure condition and the parallel condition; however, the 4-dimensional normal vectors of the tetrahedra in the $\Delta_3$ do not exist. Therefore, these saddle points give rise to Lorentzian $SO(1,3)$ bivector geometry.

\section{Conclusion}

We have developed our saddle-point finder to find the complex saddle points for any given action. Applying the saddle-point finder to two examples in the spin foam model, we find the complex saddle points and estimate their contributions to the partition function. Finding these saddle points would help not only the asymptotic analysis of the analytically continued spin foam model but also the Lefschetz thimble Monte Carlo computation in the regime of small $j$, because in this regime, the non-perturbative contribution due to the complex saddle points is non-negligible.

In the example of the $\Delta_3$ spin foam model, all of the saddle points we have found do not correspond to simplicial geometry. This result enforced the conclusion in \cite{Han:2018fmu,han2021complex}, i.e., the classical limit of the spin foam model should be taken in the limit with large-$j$ but small deficit angles.

In this paper, we use the information of the simplicial geometry to help us to narrow down the region to find the complex saddle points. In future works, instead of using the physical information, we would like to employ certain optimization algorithm in the pre-treatment stage to automatically find the proper region to be the workplace for our finder. This optimization will improve our saddle-point finder to be a "black-box" that is applicable to other physical system other than LQG.

\acknowledgments
YW is supported by NSFC grant No. 11875109, General Program of Science and Technology of Shanghai No. 21ZR1406700, and Shanghai Municipal Science and Technology Major Project (Grant No.2019SHZDZX01). YW is also grateful to the Perimeter Institute for Theoretical Physics for hospitality during his visit, where the work is finalized. ZH thanks Hongguang Liu for the useful discussions. 

\appendix
\section{Saddle points of single 4-simplex spin foam model}
\begin{table}[h]
	\centering\caption{The values of $j_{ab}$ at $s_1$.}\label{tab:facearea1}
	\small
	\setlength{\tabcolsep}{0.8mm}
	\begin{tabular}{|c|c|c|c|c|}
    \hline
		\diagbox{\small{a}}{ $j_{ab}$}{\small{b}}&2&3&4&5\\
		\hline
		1&$4.947+3.013\times10^{-4}\ii$&$4.946+2.766\times10^{-4}\ii$&$4.948+3.358\times10^{-4}\ii$&$4.951+4.880\times10^{-4}\ii$\\
		\hline
		2&\diagbox{}{}&$2.018+1.267\times10^{-3}\ii$&$2.019+1.383\times10^{-3}\ii$&$2.021+1.490\times10^{-3}\ii$\\
		\hline
		3&\diagbox{}{}&\diagbox{}{}&$2.019+1.341\times10^{-3}\ii$&$2.021+1.448\times10^{-3}\ii$\\
		\hline
		4&\diagbox{}{}&\diagbox{}{}&\diagbox{}{}&$2.0198+1.386\times10^{-3}\ii$\\
		\hline
	\end{tabular}
\end{table} 

\begin{table}[h]
	\centering\caption{The values of $g_a$ at $s_1$.}\label{tab:ga1}
	\scriptsize
	\setlength{\tabcolsep}{0.5mm}
	\begin{tabular}{|c|c|}
		\hline
		\small{a}&$g_a$\\
		\hline
		$1$ &$\left(\begin{matrix}
		1&0\\	0&1
		\end{matrix}\right)$\\
		\hline
		$2$ &
		$
		\left(
        \begin{array}{cc}
        1.793\times 10^{-2}+\left(4.688\times 10^{-4}\right) \ii & 6.919\times 10^{-5}+\left(9.997\times10^{-1}\right) \ii \\
        -5.331\times 10^{-5}+1\ii & 1.832\times 10^{-2}+\left(4.071\times 10^{-4}\right) \ii \\
        \end{array}
        \right)
        $\\
        \hline
        $3$ &
        $\left(
        \begin{array}{cc}
        1.827\times 10^{-2}+\left(6.071\times 10^{-4}\right) \ii & 9.425\times 10^{-1}-\left(3.332\times10^{-1}\right) \ii \\
        -9.428\times 10^{-1}-\left(3.333\times 10^{-1}\right) \ii & 1.797\times10^{-2}+\left(4.961\times 10^{-4}\right) \ii \\
        \end{array}
        \right)$\\
		\hline
		$4$ &
		$
		\left(
        \begin{array}{cc}
        1.814\times 10^{-2}+\left(8.171\times 10^{-1}\right) \ii & -4.712\times10^{-1}-\left(3.332\times 10^{-1}\right) \ii \\
        4.714\times 10^{-1}-\left(3.333\times 10^{-1}\right) \ii & 1.809\times 10^{-2}-\left(8.158\times10^{-1}\right) \ii \\
        \end{array}
        \right)
		$\\
		\hline
		$5$ &
		$
		\left(
        \begin{array}{cc}
        1.811\times 10^{-2}-\left(8.156\times 10^{-1}\right) \ii & -4.713\times10^{-1}-\left(3.332\times 10^{-1}\right) \ii \\
        4.713\times 10^{-1}-\left(3.333\times 10^{-1}\right) \ii & 1.81\times 10^{-2}+\left(8.171\times10^{-1}\right) \ii \\
        \end{array}
        \right)
		$\\
		\hline
	\end{tabular}
\end{table} 

\begin{table}[h]
	\centering\caption{The values of $g_a^\dagger$ at $s_1$.}\label{tab:cga1}
	\scriptsize
	\setlength{\tabcolsep}{0.5mm}
	\begin{tabular}{|c|c|}
		\hline
		\small{a}&$g_a^\dagger$\\
		\hline
		$1$ &$\left(\begin{matrix}
		1&0\\	0&1
		\end{matrix}\right)$\\
		\hline
		$2$ &
		$
		\left(
        \begin{array}{cc}
        1.832\times 10^{-2}+\left(4.071\times 10^{-4}\right) \ii & -6.919\times10^{-5}-\left(9.997\times 10^{-1}\right) \ii \\
        5.331\times 10^{-5}-1\ii & 1.793\times 10^{-2}+\left(4.688\times 10^{-4}\right) \ii \\
        \end{array}
        \right)
        $\\
        \hline
        $3$ &
        $
        \left(
        \begin{array}{cc}
        1.797\times 10^{-2}+\left(4.961\times 10^{-4}\right) \ii & -9.425\times10^{-1}+\left(3.332\times 10^{-1}\right) \ii \\
        9.428\times 10^{-1}+\left(3.333\times 10^{-1}\right) \ii & 1.827\times 10^{-2}+\left(6.071\times10^{-4}\right) \ii \\
        \end{array}
        \right)
        $\\
		\hline
		$4$ &
		$
		\left(
        \begin{array}{cc}
        1.809\times 10^{-2}-\left(8.158\times 10^{-1}\right) \ii & 4.712\times 10^{-1}+\left(3.332\times 10^{-1}\right) \ii \\
        -4.714\times 10^{-1}+\left(3.333\times 10^{-1}\right) \ii & 1.814\times 10^{-2}+\left(8.171\times 10^{-1}\right) \ii \\
        \end{array}
        \right)
		$\\
		\hline
		$5$ &
		$
		\left(
        \begin{array}{cc}
        1.81\times 10^{-2}+\left(8.171\times 10^{-1}\right) \ii & 4.713\times 10^{-1}+\left(3.332\times 10^{-1}\right) \ii \\
        -4.713\times 10^{-1}+\left(3.333\times 10^{-1}\right) \ii & 1.811\times 10^{-2}-\left(8.156\times 10^{-1}\right) \ii \\
        \end{array}
        \right)
		$\\
		\hline
	\end{tabular}
\end{table}

\begin{table}[htbp]
  \centering\caption{Values of $z_{ab}$ at $s_1$.}\label{tab:zv1}
\footnotesize
\resizebox{\textwidth}{15mm}{
\begin{tabular}{|c|c|c|c|c|c|}
  \hline
  \diagbox{\small{a}}{$\ket{z_{ab}}$ }{\small{b}}&1&2&3&4&5\\
  \hline
  1&\diagbox{}{}&(1,1)&(1,-0.3333+0.9428i)&(1,-0.1835-0.2595i)&(1,-1.816-2.569i)\\
  \hline
  2&(1,1)&\diagbox{}{}&(1,0.8507-0.5636i)&(1,1.568+0.5511i)&(1,0.5603+0.1737i)\\
  \hline
  3&(1,-0.3333+0.9428i)&(1,0.8507-0.5636i)&\diagbox{}{}&(1,-0.067+1.704i)&(1,-0.001082+0.6021i)\\
  \hline
  4&(1,-0.1835-0.2595i)&(1,1.568+0.5511i)&(1,-0.067+1.704i)&\diagbox{}{}&(1,-0.01705-0.006849i)\\
  \hline
  5&(1,-1.816-2.569i)&(1,0.5603+0.1737i)&(1,-0.001082+0.6021i)&(1,-0.01705-0.006849i)&\diagbox{}{}\\
  \hline
\end{tabular}
}
\end{table}

\begin{table}[htbp]
  \centering\caption{Values of Conjugate $z_{ab}$ at $s_1$.}\label{tab:czv1}
\footnotesize
\resizebox{\textwidth}{15mm}{
\begin{tabular}{|c|c|c|c|c|c|}
  \hline
  \diagbox{\small{a}}{$\bra{z_{ab}}$ }{\small{b}}&1&2&3&4&5\\
  \hline
  1&\diagbox{}{}&(1,0.9997-0.0005081i)&(1,-0.3328-0.9426i)&(1,-0.1834+0.2593i)&(1,-1.817+2.569i)\\
  \hline
  2&(1.,0.9997-0.0005081i)&\diagbox{}{}&(1,0.8145+0.6141i)&(1,1.633-0.6519i)&(1,0.5211-0.1853i)\\
  \hline
  3&(1,-0.3328-0.9426i)&(1,0.8145+0.6141i)&\diagbox{}{}&(1,0.003138-1.808i)&(1,0.02253-0.568i)\\
  \hline
  4&(1,-0.1834+0.2593i)&(1,1.633-0.6519i)&(1,0.003138-1.808i)&\diagbox{}{}&(1,0.0007018-0.01837i)\\
  \hline
  5&(1,-1.817+2.569i)&(1,0.5211-0.1853i)&(1,0.02253-0.568i)&(1,0.0007018-0.01837i)&\diagbox{}{}\\
  \hline
\end{tabular}
}
\end{table}

\begin{table}[h]
	\centering\caption{The values of $j_{ab}$ at  $s_2$.}\label{tab:facearea2}
	\small
	\setlength{\tabcolsep}{0.8mm}
	\begin{tabular}{|c|c|c|c|c|}
    \hline
		\diagbox{\small{a}}{ $j_{ab}$}{\small{b}}&2&3&4&5\\
		\hline
		1&4.976-0.07024i&4.976-0.07097i&4.977-0.06961i&4.977-0.06565i\\
		\hline
		2&\diagbox{}{}&2.004+0.02366i&2.004+0.02595i&,2.005+0.0285i\\
		\hline
		3&\diagbox{}{}&\diagbox{}{}&2.004+0.02514i&2.005+0.0277i\\
		\hline
		4&\diagbox{}{}&\diagbox{}{}&\diagbox{}{}&2.005+0.02678i\\
		\hline
	\end{tabular}
\end{table} 

\begin{table}[h]
	\centering\caption{The values of $g_a$ at $s_2$.}\label{tab:ga2}
	\scriptsize
	\setlength{\tabcolsep}{0.5mm}
	\begin{tabular}{|c|c|}
		\hline
		\small{a}&$g_a$\\
		\hline
		$1$ &$\left(\begin{matrix}
		1&0\\	0&1
		\end{matrix}\right)$\\
		\hline
		$2$ &
		$
		\left(
        \begin{array}{cc}
        -8.793\times 10^{-2}-\left(1.918\times 10^{-1}\right) \ii & 1.683\times 10^{-2}+1.015 \ii \\
        1.632\times 10^{-2}+1.014 \ii & -8.766\times 10^{-2}-\left(1.913\times 10^{-1}\right) \ii \\
        \end{array}
        \right)
        $\\
        \hline
        $3$ &
        $
        \left(
        \begin{array}{cc}
        -8.788\times 10^{-2}-\left(1.913\times 10^{-1}\right) \ii & 9.508\times 10^{-1}-\left(3.539\times 10^{-1}\right) \ii \\
        -9.621\times 10^{-1}-\left(3.228\times 10^{-1}\right) \ii & -8.785\times 10^{-2}-\left(1.918\times 10^{-1}\right) \ii \\
        \end{array}
        \right)
        $\\
		\hline
		$4$ &
		$
		\left(
        \begin{array}{cc}
        -7.448\times 10^{-2}+\left(6.368\times 10^{-1}\right) \ii & -4.838\times 10^{-1}-\left(3.3\times 10^{-1}\right) \ii \\
        4.728\times 10^{-1}-\left(3.459\times 10^{-1}\right) \ii & -1.015\times 10^{-1}-1.02 \ii \\
        \end{array}
        \right)
		$\\
		\hline
		$5$ &
		$
		\left(
        \begin{array}{cc}
        -1.018\times 10^{-1}-1.02 \ii & -4.838\times 10^{-1}-\left(3.303\times 10^{-1}\right) \ii \\
        4.727\times 10^{-1}-\left(3.46\times 10^{-1}\right) \ii & -7.463\times 10^{-2}+\left(6.368\times 10^{-1}\right) \ii \\
        \end{array}
        \right)
		$\\
		\hline
	\end{tabular}
\end{table} 

\begin{table}[h]
	\centering\caption{ The values of $g_a^\dagger$ at $s_2$.}\label{tab:cga2}
	\scriptsize
	\setlength{\tabcolsep}{0.5mm}
	\begin{tabular}{|c|c|}
		\hline
		\small{a}&$g_a^\dagger$\\
		\hline
		$1$ &$\left(\begin{matrix}
		1&0\\	0&1
		\end{matrix}\right)$\\
		\hline
		$2$ &
		$
		\left(
        \begin{array}{cc}
        1.588\times 10^{-1}+\left(1.664\times 10^{-1}\right) \ii & -2.647\times 10^{-2}-1.002 \ii \\
        -2.619\times 10^{-2}-1.002 \ii & 1.586\times 10^{-1}+\left(1.659\times 10^{-1}\right) \ii \\
        \end{array}
        \right)
        $\\
        \hline
        $3$ &
        $
        \left(
        \begin{array}{cc}
        1.585\times 10^{-1}+\left(1.661\times 10^{-1}\right) \ii & -9.354\times 10^{-1}+\left(3.588\times 10^{-1}\right) \ii \\
        9.531\times 10^{-1}+\left(3.092\times 10^{-1}\right) \ii & 1.585\times 10^{-1}+\left(1.665\times 10^{-1}\right) \ii \\
        \end{array}
        \right)
        $\\
		\hline
		$4$ &
		$
		\left(
        \begin{array}{cc}
        1.37\times 10^{-1}-\left(6.516\times 10^{-1}\right) \ii & 4.81\times 10^{-1}+\left(3.213\times 10^{-1}\right) \ii \\
        -4.634\times 10^{-1}+\left(3.462\times 10^{-1}\right) \ii & 1.798\times 10^{-1}+\left(9.842\times 10^{-1}\right) \ii \\
        \end{array}
        \right)
		$\\
		\hline
		$5$ &
		$
		\left(
        \begin{array}{cc}
        1.798\times 10^{-1}+\left(9.84\times 10^{-1}\right) \ii & 4.81\times 10^{-1}+\left(3.214\times 10^{-1}\right) i \\
        -4.635\times 10^{-1}+\left(3.463\times 10^{-1}\right) \ii & 1.369\times 10^{-1}-\left(6.516\times 10^{-1}\right) \ii \\
        \end{array}
        \right)
		$\\
		\hline
	\end{tabular}
\end{table}

\begin{table}[htbp]
  \centering\caption{Values of $z_{ab}$ at $s_2$.}\label{tab:zv2}
\footnotesize
\resizebox{\textwidth}{15mm}{
\begin{tabular}{|c|c|c|c|c|c|}
  \hline
  \diagbox{\small{a}}{$\ket{z_{ab}}$ }{\small{b}}&1&2&3&4&5\\
  \hline
  1&\diagbox{}{}&(1,1-0.00007656i)&(1,-0.3334+0.9427i)&(1,-0.1835-0.2595i)&(1,-1.816-2.569i)\\
  \hline
  2&(1,1-0.00007656i)&\diagbox{}{}&(1,1.044-0.4492i)&(1,1.258+0.4235i)&(1,0.6495+0.06107i)\\
  \hline
  3&(1,1.044-0.4492i)&(1,0.8507-0.5636i)&\diagbox{}{}&(1,-0.3736+1.486i)&(1,-0.01186+0.753i)\\
  \hline
  4&(1,1.258+0.4235i)&(1,1.568+0.5511i)&(1,-0.3736+1.486i)&\diagbox{}{}&(1,-0.113-0.04805i)\\
  \hline
  5&(1,-1.816-2.569i)&(1,0.6495+0.06107i)&(1,-0.01186+0.753i)&(1,-0.113-0.04805i)&\diagbox{}{}\\
  \hline
\end{tabular}
}
\end{table}

\begin{table}[htbp]
  \centering\caption{Values of Conjugate $z_{ab}$ at $s_2$.}\label{tab:czv2}
\footnotesize
\resizebox{\textwidth}{15mm}{
\begin{tabular}{|c|c|c|c|c|c|}
  \hline
  \diagbox{\small{a}}{$\bra{z_{ab}}$ }{\small{b}}&1&2&3&4&5\\
  \hline
  1&\diagbox{}{}&(1,1.001-0.0006011i)&(1,-0.3335-0.9421i)&(1,-0.1832+0.2596i)&(1,-1.816+2.569i)\\
  \hline
  2&(1,1.001-0.0006011i)&\diagbox{}{}&(1,0.8086+0.3475i)&(1,1.527-0.1443i)&(1,0.7141-0.2405i)\\
  \hline
  3&(1,-0.3335-0.9421i)&(1,0.8086+0.3475i)&\diagbox{}{}&(1,-0.0209-1.327i)&(1,-0.1588-0.6324i)\\
  \hline
  4&(1,-0.1832+0.2596i)&(1,1.527-0.1443i)&(1,-0.0209-1.327i)&\diagbox{}{}&(1,-0.007496+0.1226i)\\
  \hline
  5&(1,-1.816+2.569i)&(1,0.7141-0.2405i)&(1,-0.1588-0.6324i)&(1,-0.007496+0.1226i)&\diagbox{}{}\\
  \hline
\end{tabular}
}
\end{table}

\section{The saddle point $X_0$}

\begin{table}[h]
	\centering\caption{The table of $g^0_{ab}$ }\label{tab:g0a1}
	\footnotesize
	\setlength{\tabcolsep}{0.8mm}
	\resizebox{\textwidth}{26mm}{
	\begin{tabular}{|c|c|c|c|}
		\hline
		\small{\diagbox{b}{$g_{ab}^0$}{a}}&6&4&2\\
		\hline
		$1$ &$\left(\begin{matrix}
		0.9553&-0.2955\ii\\	-0.2955\ii&0.9553
		\end{matrix}\right)$
		&
		$\left(\begin{matrix}
		-0.3900+0.6198\ii&-0.1417-0.6688\ii\\	0.1401-0.6650\ii&-0.3888-0.6193\ii
		\end{matrix}\right)$
		&
		$\left(\begin{matrix}
		0.7052+0.04132\ii&-0.1336-0.3004\ii\\	
		2.173-1.447\ii&0.3506-0.6723\ii
		\end{matrix}\right)$
		\\
		\hline
		$2$ &$\left(\begin{matrix}
		0.4515+0.5054\ii&-1.042-0.4024\ii\\	0.4792-0.2790\ii&0.4896-0.3313\ii
		\end{matrix}\right)$
		&
		$\left(\begin{matrix}
		0.1784-0.6465\ii&0.4982-0.5489\ii\\	-0.4970-0.5489\ii&0.1799+0.6483\ii
		\end{matrix}\right)$
		&
		\diagbox{}{}
		\\
        \hline
		$3$ &$\left(\begin{matrix}
		0.8343-0.1999\ii&0.6464+0.7435\ii\\	-0.3138+0.2731\ii&0.6888+0.09706\ii
		\end{matrix}\right)$
		&
		$\left(\begin{matrix}
		0.2856+0.1372\ii&0.1484-0.9387\ii\\	-0.1479-0.9346\ii&0.2869-0.1373\ii
		\end{matrix}\right)$
		&
		$\left(\begin{matrix}
		0.9553&-0.2955\ii\\	-0.2955\ii&0.9553
		\end{matrix}\right)$
		\\
		\hline
		$4$ &$\left(\begin{matrix}
		0.6724+0.06192\ii&-0.2002-0.5181\ii\\  0.1395-0.5346\ii&1.030-0.04314\ii
		\end{matrix}\right)$
		&
		\diagbox{}{}
		&
		$\left(\begin{matrix}
		1.2080+0.2345\ii&0.8229-0.2524\ii\\	
		-1.094+0.1394\ii&0.1679+0.2911\ii
		\end{matrix}\right)$
		\\
		\hline
		$5$ &$\left(\begin{matrix}
		0.1820-0.2099\ii&0.08676-1.274\ii\\  -0.01552-0.7019\ii&0.3588+0.1879\ii
		\end{matrix}\right)$
		&
		$\left(\begin{matrix}
		0.9553&-0.2955\ii\\  -0.2955\ii&0.9553
		\end{matrix}\right)$
		&
		$\left(\begin{matrix}
		0.7194-0.2650\ii&-0.1255-0.06036\ii\\  
		0.7314+2.2003\ii&1.419+0.07753\ii
		\end{matrix}\right)$
		\\
		\hline
		$6$ &\diagbox{}{}
		&
		$\left(\begin{matrix}
		0.4773-0.06975\ii&0.2255-0.8464\ii\\  -0.2249-0.8453\ii&0.4797+0.06956\ii
		\end{matrix}\right)$
		&
		$\left(\begin{matrix}
		2.096-0.2671\ii&-0.3216-0.5575\ii\\  
		0.6308+0.1225\ii&0.4297-0.1318\ii
		\end{matrix}\right)$
		\\
		\hline
	\end{tabular}
	}
\end{table}

\begin{table}[htbp]
  \centering\caption{The table of $z_{6ab}^0$.}\label{tab:zv6}
\footnotesize
\resizebox{\textwidth}{15mm}{
\begin{tabular}{|c|c|c|c|c|c|}
  \hline
  \diagbox{\small{a}}{$z_{6ab}^0$ }{\small{b}}&1&2&3&4&5\\
  \hline
  1&\diagbox{}{}&(1,-1.615+1.503i)&(1,-0.2227-0.5883i)&(1,-0.3093i)&(1,-0.1173-0.02850i)\\
  \hline
  2&(1,-1.615+1.503i)&\diagbox{}{}&(1,0.5763-0.03732i)&(1,0.5401-0.2764i)&(1,0.7037-0.4995i)\\
  \hline
  3&(1,-0.2227-0.5883i)&(1,0.5763-0.03732i)&\diagbox{}{}&(1,0.3919+0.4517i)&(1,-0.03505+0.2601i)\\
  \hline
  4&(1,-0.3093i)&(1,0.5401-0.2764i)&(1,0.3919+0.4517i)&\diagbox{}{}&(1,-0.1737+0.1311i)\\
  \hline
  5&(1,-0.1173-0.02850i)&(1,0.7037-0.4995i)&(1,-0.03505+0.2601i)&(1,-0.1737+0.1311i)&\diagbox{}{}\\
  \hline
\end{tabular}
}
\end{table}

\begin{table}[htbp]
  \centering\caption{The table of $z_{4ab}^0$.}\label{tab:zv4}
\footnotesize
\resizebox{\textwidth}{15mm}{
\begin{tabular}{|c|c|c|c|c|c|}
  \hline
  \diagbox{\small{a}}{$z_{4ab}^0$ }{\small{b}}&1&2&3&5&6\\
  \hline
  1&\diagbox{}{}&(1,-0.1445+0.9792i)&(1,-0.2163+0.5057i)&(1,0.3001-0.5271i)&(1,0.7687-0.7459i)\\
  \hline
  2&(1,-0.1445+0.9792i)&\diagbox{}{}&(1,-0.2137+0.1556i)&(1,0.3251-0.1581i)&(1,0.9837+0.4936i)\\
  \hline
  3&(1,-0.2163+0.5057i)&(1,-0.2137+0.1556i)&\diagbox{}{}&(1,0.05290+0.003738i)&(1,0.09562+0.3193i)\\
  \hline
  5&(1,0.3001-0.5271i)&(1,0.3251-0.1581i)&(1,0.05290+0.003738i)&\diagbox{}{}&(1,-0.3093i)\\
  \hline
  6&(1,0.7687-0.7459i)&(1,0.9837+0.4936i)&(1,0.09562+0.3193i)&(1,-0.3093i)&\diagbox{}{}\\
  \hline
\end{tabular}
}
\end{table}

\begin{table}[htbp]
  \centering\caption{The table of $z_{2ab}^0$.}\label{tab:zv22}
\footnotesize
\resizebox{\textwidth}{15mm}{
\begin{tabular}{|c|c|c|c|c|c|}
  \hline
  \diagbox{\small{a}}{$z_{2ab}^0$ }{\small{b}}&1&3&4&5&6\\
  \hline
  1&\diagbox{}{}&(1,-2.928+2.087i)&(1,-3.280+2.139i)&(1,-3.498+1.795i)&(1,-2.962+1.585i)\\
  \hline
  3&(1,-2.928+2.087i)&\diagbox{}{}&(1,-1.442+1.530i)&(1,-0.5935+1.644i)&(1,0.6318+3.250i)\\
  \hline
  4&(1,-3.280+2.139i)&(1,-1.442+1.530i)&\diagbox{}{}&(1,-0.5022+1.623i)&(1,1.268+3.028i)\\
  \hline
  5&(1,-3.498+1.795i)&(1,-0.5935+1.644i)&(1,-0.5022+1.623i)&\diagbox{}{}&(1,2.085+3.439i)\\
  \hline
  6&(1,-2.962+1.585i)&(1,0.6318+3.250i)&(1,1.268+3.028i)&(1,2.085+3.439)&\diagbox{}{}\\
  \hline
\end{tabular}
}
\end{table}

\section{Parity flipping}
In section 6, $g_{64}$, $g_{42}$, and $g_{26}$ are parameterized as in \eqref{eq:sup}. 
Considering a parity flip at $4$-simplex $6$, the saddle point values of $g_{64}$ and $g_{46}$ are $({g^{0}}^\dagger_{64})^{-1}$ and $g^0_{46}$. 
In fact, there is no upper triangular matrix $T$, such that $g^0_{64}T=({g^{0}}^\dagger_{64})^{-1}$. 
But the parameterization used in section 6 is still compatible with the parity flipping. By $\Su$ gauge, $g_{64}=({g^{0}}^\dagger_{64})^{-1},\, g_{46}=g^0_{46}$ is equivalent to $g_{64}=({g^{0}}^\dagger_{64})^{-1} U,\, g_{46}=g^0_{46} U$, where $U\in\Su$. One can always find a upper triangular matrix $T$ such that $g^0_{64}T=({g^{0}}^\dagger_{64})^{-1} U$. Thus, the parity flipped saddle point with $g_{64}=({g^{0}}^\dagger_{64})^{-1} U,\, g_{46}=g^0_{46} U$ can be expressed in our parameterization. Explicitly, solving the equation
\[
\left({g^{0}}^\dagger_{64}g^{0}_{64}\right)^{-1}\cdot\left(\left({g^{0}}^\dagger_{64}g^{0}_{64}\right)^{-1}\right)^\dagger=T\cdot T^\dagger,
\]
results in $T$, and 
\[
U={g^{0}}^\dagger_{64}g^{0}_{64}T.
\]
The parity flip on $4$-simplex $4$ or on $4$-simplex $2$ can be treated similarly.

\bibliographystyle{unsrt}
\bibliography{graviton.bib}


\end{document}